\documentclass[aps,prx,twocolumn,amsmath,amssymb,superscriptaddress,10pt,nobibnotes]{revtex4-2}
\usepackage{preamble}

\begin{document}


\title{Model-free estimation of the Cramér-Rao bound\texorpdfstring{\\}{}for deep-learning microscopy in complex media}
\author{Ilya Starshynov}
\thanks{These authors contributed equally to this work.}
\affiliation{School of Physics and Astronomy, University of Glasgow, Glasgow, G12 8QQ, United Kingdom}
\author{Maximilian Weimar}
\thanks{These authors contributed equally to this work.}
\affiliation{Institute for Theoretical Physics, Vienna University of Technology (TU Wien), 1040 Vienna, Austria}
\author{Lukas M.\ Rachbauer}
\affiliation{Institute for Theoretical Physics, Vienna University of Technology (TU Wien), 1040 Vienna, Austria}
\author{Günther Hackl}
\affiliation{Institute for Theoretical Physics, Vienna University of Technology (TU Wien), 1040 Vienna, Austria}
\author{Daniele Faccio}
\affiliation{School of Physics and Astronomy, University of Glasgow, Glasgow, G12 8QQ, United Kingdom}
\author{Stefan Rotter}
\affiliation{Institute for Theoretical Physics, Vienna University of Technology (TU Wien), 1040 Vienna, Austria}
\author{Dorian Bouchet}
\altaffiliation{\href{mailto:dorian.bouchet@univ-grenoble-alpes.fr}{dorian.bouchet@univ-grenoble-alpes.fr}}
\affiliation{Univ.\ Grenoble Alpes, CNRS, LIPhy, 38000 Grenoble, France}


\begin{abstract}
Artificial neural networks have become important tools to harness the complexity of disordered or random photonic systems. Recent applications include the recovery of information from light that has been scrambled during propagation through a complex scattering medium, especially in the challenging case where the deterministic input–output transmission matrix cannot be measured. This naturally raises the question of what the limit is that information theory imposes on this recovery process, and whether neural networks can actually reach this limit. To answer these questions, we introduce a model-free approach to calculate the Cramér-Rao bound, which sets the ultimate precision limit at which artificial neural networks can operate. As an example, we apply this approach in a proof-of-principle experiment using laser light propagating through a disordered medium, evidencing that a convolutional network approaches the ultimate precision limit in the challenging task of localizing a reflective target hidden behind a dynamically-fluctuating scattering medium. The model-free method introduced here is generally applicable to benchmark the performance of any deep-learning microscope, to drive algorithmic developments and to push the precision of metrology and imaging techniques to their ultimate limit.

\end{abstract}

\maketitle


Complexity and related chaotic processes lie at the basis of many physical phenomena. Traditionally, complexity has hindered the capability to predict the evolution of physical systems in various research fields ranging from biophotonics~\cite{moen_deep_2019,tian_deep_2021} to quantum optics~\cite{gebhart_learning_2023}. However, over the past few years, many of these fields have seen remarkable progress as a result of data-driven models and machine learning approaches, that are capable of harnessing the complexity of physical dynamics with a surprising effectiveness \cite{tang_introduction_2020}. In essence, these data-driven models are processing the physical information from the training data and building a representative statistical model from this data. The question of how exactly this information is maintained and distilled by an artificial neural network (ANN) then naturally arises, especially for applications in photonics that require the reconstruction of images \cite{barbastathis_use_2019} or the precise estimation of physical observables \cite{zuo_deep_2022}. 

Light propagation through a complex medium is a typical example of a research area that has seen major advances thanks to data-driven models \cite{gigan_roadmap_2022}. While the propagation of light is ruled by simple laws in homogeneous media, retrieving information through complex scattering media is a critical challenge, as light typically undergoes a number of unknown scattering and absorption events during propagation~\cite{yoon_deep_2020}. Several imaging techniques that were originally developed to address the challenges associated with complex light scattering are based on the insight that the measured output random patterns (referred to as speckle patterns) are the deterministic results of millions of scattering events \cite{mosk_controlling_2012}. Notably, a popular approach consists in measuring experimentally the deterministic relation between object and image planes, which is conveniently done in a scattering matrix formalism \cite{popoff_measuring_2010,conkey_high-speed_2012,yu_measuring_2013}; the knowledge of this matrix can then be used to effectively transform the scattering medium into a simple optical element \cite{popoff_image_2010,choi_overcoming_2011}. Nevertheless, in many cases of interest, including those involving dynamical scattering media that change in time, this deterministic relation cannot be measured without the use of a guidestar~\cite{horstmeyer_guidestar-assisted_2015}, which limits the practical applicability of the approach. 

To reconstruct images through such media considering the occurrence of random scattering events, a number of techniques have been developed such as multiphoton microscopy \cite{denk_two-photon_1990,horton_vivo_2013}, optical coherence tomography \cite{huang_optical_1991,kashani_optical_2017} and correlation imaging \cite{bertolotti_non-invasive_2012,katz_non-invasive_2014}, but these are limited to specific application scenarios. In recent years, the emergence of ANNs has been a game changer in this field; indeed, deep neural networks were shown to be able to learn to image through a known or fixed scattering medium, or even reconstruct images hidden by unseen scattering media, with a remarkable fidelity~\cite{scatter1,horisaki_learning-based_2016,li_imaging_2018,li_deep_2018,sun_efficient_2018,scatter2,fiber1,lyu_learning-based_2019,fiber4,fiber2,scatter6,li_displacement-agnostic_2021,scatter3,fiber3,bai_all-optical_2023,abdulaziz_robust_2023,Huang:24}.

\begin{figure*}[t]
    \begin{center}
		\includegraphics[width=\textwidth]{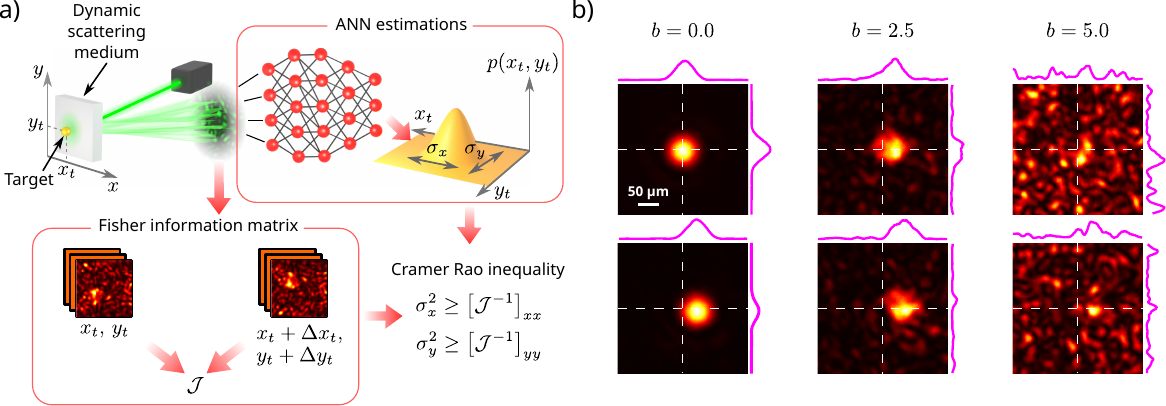}
	\end{center}
    \caption{Principle of the experiment. (a) A reflective target is placed behind a dynamic scattering sample. Our goal is to infer its position from the reflected coherent light using different ANNs, and compare the precision of the ANNs to the ultimate limit calculated using Fisher information theory. (b) Examples of pairs of measured images with two adjacent target positions (see the top and bottom rows) at different scattering strengths (see the three columns labeled by the optical thickness $b=L/\ell$). In the case of a strongly scattering sample ($b=5$), the precise target position cannot be easily estimated from such images due to random light scattering occurring within the dynamic medium.}   
	\label{fig_principle}
\end{figure*}

In order to benchmark the performance of these techniques, an insightful strategy consists in studying fundamental limits imposed by physical laws on the quality of the reconstructed images. One such limit is the Abbe limit, which describes the ultimate resolution achievable with an imaging system. However, while this limit is well-defined in homogeneous media, it cannot be used when imaging through scattering media. Indeed, multiple scattering effects alter light fields in a complex way, not only affecting the resolution but also impacting its contrast, possibly with the apparition of artifacts. Moreover, resolution is not a relevant metric to describe the performance of most computational imaging techniques, including those based on ANNs, since they often break the Abbe limit by including prior information about the object in the reconstruction procedure. 

Instead of image resolution, we propose here to use a criterion that overcomes these shortcomings and sets a quantitative benchmark to assess the ability of ANNs to extract information from physical measurements. This criterion, which is commonly employed in optical metrology~\cite{giovannetti_advances_2011,polino_photonic_2020} and which is generally applicable to any physical system, is implemented and demonstrated here for the specific case of light propagation in complex scattering media. The approach relies on the assumption that the required information is not a global image in itself, but specific features of interest that could potentially be extracted from this image, such as the size or the position of an object \cite{cohen_subwavelength_2011,del_hougne_precise_2018,del_hougne_robust_2020,jauregui-sanchez_tracking_2022}. In this framework, the relevant physical limit is the Cramér-Rao bound, which sets the ultimate precision with which parameters of interest can be estimated \cite{van_trees_detection_2013}. However, calculating this bound is a challenging task in complex systems. Indeed, the expression of the Cramér-Rao bound is based on the probability density function describing the data statistics, the analytical expression of which is typically unknown. 

Here, we overcome this hurdle by introducing an approach to calculate the Cramér-Rao bound solely from experimental data, even when a physical model describing the data is not available. We then use this benchmark to assess the performance of different types of ANNs that we have trained to estimate the position of an object through a dynamic scattering sample. For this task, our analysis demonstrates that a convolutional neural network approaches the limit set by the Cramér-Rao bound. In our experiments, we consider the canonical case of a target object hidden behind a random scattering medium, as shown in \ffig{fig_principle}{a}. We illuminate the medium with a laser, which gets transmitted to the hidden object as a random speckle pattern of light. The retro-reflected light passes once more back through the random medium before being collected by a camera (see \methods{} for a detailed description of the optical setup). Our aim is to precisely estimate the position $\theta=(x_t,y_t)$ of this object from a single intensity frame collected by the camera. In practice, we use as an object a reflecting target displayed by a digital micromirror device (DMD). This target is hidden behind a dynamic random scattering sample composed of a suspension of TiO$_2$ particles in glycerol, which is being pumped through a flow cuvette. The decorrelation time is such that two successive frames are uncorrelated (see \methods{}). In this configuration, we can vary the concentration of TiO$_2$ particles to tune the scattering mean free path $\ell$ and control the optical thickness $b=L/\ell$ of the medium, where $L$ is the optical path in the cuvette. In order to study different scattering regimes, we vary the optical thickness $b$ from 1.7 to 5, that we also compare to the case in which light propagates in free space ($b=0$). While the target can be easily localized in free space, the occurrence of random scattering events results in the generation of speckle patterns that completely conceal the target in the case of large optical thicknesses [see \ffig{fig_principle}{b} and \SI{2}].


\section*{Results}

\subsection*{Estimation of the Cramér-Rao bound}

\begin{figure*}[t]
     \begin{center}
		\includegraphics[width=0.9\textwidth]{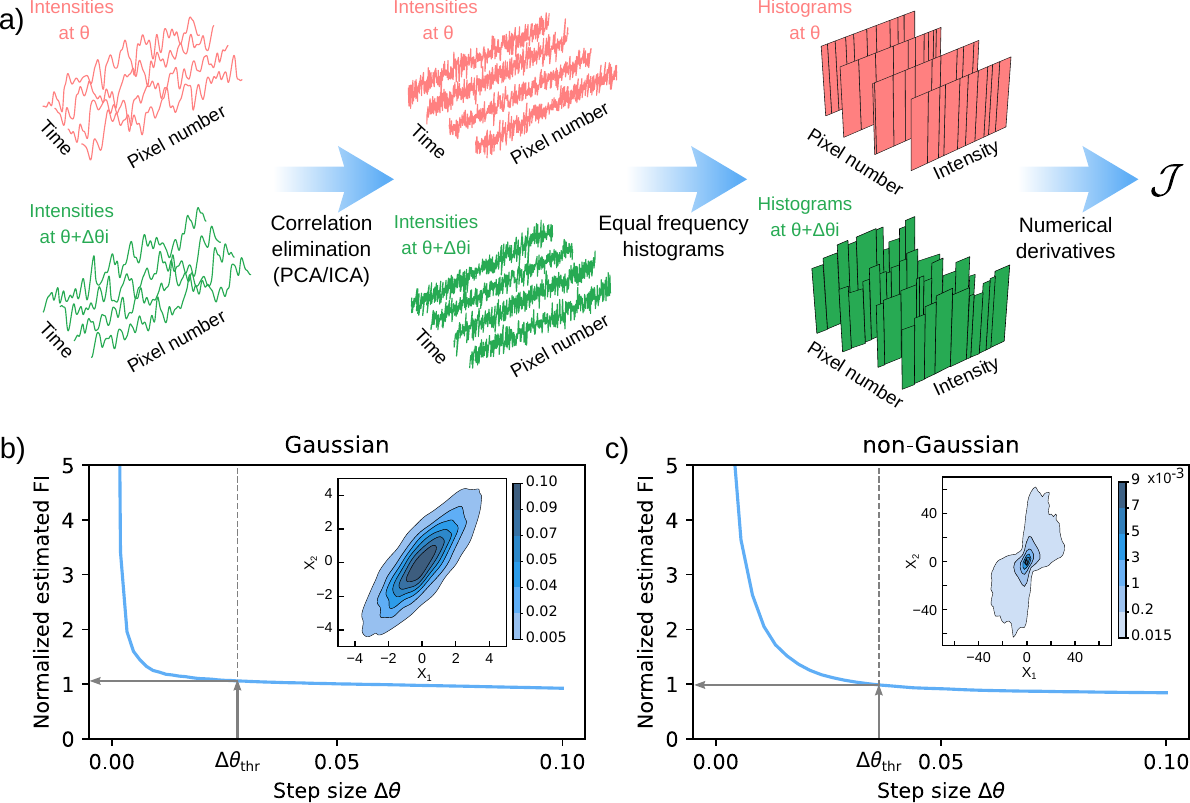}
	  \end{center}
     \caption{Demonstration of the Fisher information estimation procedure using numerically-generated data. (a)~Representation of the procedure used to evaluate the Fisher information. Images (raw data) are measured as a function of time, i.e., for different noise realizations (left panels). Statistical independence between different image pixels is restored using either PCA for Gaussian statistics or ICA for non-Gaussian statistics (middle panels). The underlying probability distributions are then estimated using equal frequency binning (right panels). Replicating this analysis for different values of $\theta$ (represented in red and green) allows us to estimate the Fisher information with a finite-difference scheme. (b)~Numerical demonstration of the procedure for data with correlated Gaussian statistics (following the distribution shown in the inset). The blue curve depicts the estimated Fisher information divided by the true Fisher information, as a function of the step size of the finite-difference scheme. A clear plateau is observed when the estimated Fisher information reaches the true Fisher information, the optimal value of the step size $\Delta\theta_{\text{thr}}$ being identified using a stability criterion based on the second derivative of this function. (c)~Analogous to (b) for data with correlated non-Gaussian statistics, demonstrating the efficiency of the method even for such complex statistics.}   
     \label{fig_fi_method}
\end{figure*}

Our first goal is to assess the ultimate localization precision achievable through dynamic scattering media. For this purpose, we employ a general framework based on statistical estimation theory~\cite{van_trees_detection_2013}. Due to statistical fluctuations arising from random interactions between the complex medium and the probe field, the connection between measured data $X$ and parameters of interest $\theta$ is intrinsically probabilistic. We can thus describe it using the probability density function (PDF) $p(X;\theta)$, which is parameterized by the vector $\theta$ containing all parameters of interest, such as the spatial coordinates $x_t$ and $y_t$ of the target. Note that this PDF is also known as the likelihood function. In this picture, measured data are represented by the random vector $X$, composed of $N$ random variables $X_k$ representing the intensity values measured by each pixel of the camera. Then, the Cram\'er-Rao inequality sets an ultimate limit on the precision of the estimated values of $\theta$. More precisely, the standard deviation $\sigma_i(\theta)$ of any unbiased estimator of the $i$-th component of $\theta$ satisfies~\cite{van_trees_detection_2013}
\begin{equation}
	\sigma_i (\theta) \geq \mathcal{C}_i(\theta) = \sqrt{\left[{\mathcal{J}}^{-1}(\theta)\right]_{ii}} ,
	\label{eq_crlb}
\end{equation} 
where $\mathcal{J}(\theta)$ is the Fisher information matrix and $\mathcal{C}_i(\theta)$ is the Cramér-Rao bound on the $i$-th component of $\theta$. The Fisher information matrix is defined by 
\begin{equation}
	\left[\mathcal{J}(\theta)\right]_{ij} =\left \langle \left[ \frac{\partial \ln p(X;\theta)}{\partial \theta_i} \right] \left[ \frac{\partial \ln p(X;\theta)}{\partial \theta_j} \right] \right \rangle ,
	\label{eq_fisher}
\end{equation}
where $\langle \cdots \rangle$ denotes the average over statistical fluctuations.

This formalism is very general, as it sets an ultimate limit on the precision of the estimated values of $\theta$ regardless of the physical origin of statistical fluctuations. Typically, the Cramér-Rao bound is calculated for shot-noise limited measurements, or for measurements corrupted by an additive Gaussian noise. Examples include single-molecule localization microscopy~\cite{ober_localization_2004,deschout_precisely_2014}, non-line-of-sight configurations~\cite{song_intelligent_2022} and static scattering media described by known scattering matrices~\cite{bouchet_influence_2020,bouchet_maximum_2021,PhysRevLett.127.233201}. In such cases, measurements follow either Gaussian or Poissonian statistics, and calculating the Cramér-Rao bound is relatively straightforward as \eq{eq_fisher} then takes a simple analytical form. In contrast, for complex scattering media that change in time, statistical fluctuations are dominated by random scattering events and no simple analytical solutions are available for \eq{eq_fisher}. Indeed, the noise statistics is unknown, and the random variables $X_k$ (which represent the measured pixel values) are correlated with each other and do not follow a simple parametric model. We thus need to rely on non-parametric estimations of the distribution, which is challenging due to the high dimension of the random variable.

In this work, we show how to overcome this difficulty by evaluating the Cramér-Rao bound solely from experimental measurements, approximating the underlying PDF from a finite number of samples [see \ffig{fig_fi_method}{a} for a graphical illustration of the approach]. First, we address the problem of correlations by applying a transformation to the data that preserves the information content but removes the dependencies between the components of $X$. A common approach, based on principal component analysis \cite{PCA} (PCA), is to estimate the covariance matrix of the data and choose a transformation that diagonalizes this matrix. This approach, however, does not guarantee statistical independence between the components. To improve on this method, we employ a so-called independent component analysis (ICA) \cite{hyvarinen2000}, which allows to construct a linear transformation $Y=AX$ that minimizes the dependencies between components of the random variable, harnessing more degrees of freedom for choosing the transformation. From the transformed data, we then estimate the underlying density function $p(Y;\theta)$ using equal frequency histograms, which allows for more stable derivative estimates as compared to histograms based on uniform bin widths. This enables us to construct an estimator of the Fisher information by approximating derivatives of $p(Y;\theta)$ with a finite difference scheme. The value of the step size $\Delta \theta$ is chosen by evaluating our estimator as a function of $\Delta \theta$ and using a stability criterion based on the second derivative of this function (see \methods{} and \SI{3.4}).

To demonstrate the performance of the approach, we test this on numerically generated data with known Fisher information. We first generate 50-dimensional multivariate Gaussian data that we decorrelate using PCA (non-Gaussianity of the data is a requirement for the ICA algorithm to converge, but PCA does guarantee independence for Gaussian statistics). We show in \ffig{fig_fi_method}{b} the estimated normalized Fisher information as a function of the step size of the finite difference scheme; a clear plateau can be identified when the normalized Fisher information reaches unity, which means that the Fisher information is correctly estimated. We then generate correlated non-Gaussian data, which we process using ICA. Even for these complex statistics, we observe again a clear plateau when the normalized Fisher information reaches unity, illustrating the broad applicability of the method.

\bigskip


\subsection*{Comparison of different ANN architectures}

We now want to assess whether the precision of ANNs can approach the Cramér-Rao bound. For this purpose, we investigate various ANN architectures to evaluate their performance in estimating the target position from measured images. We first collect a number of speckle patterns with the target located in the center of the field of view, i.e., $\theta=(0,0)$. We then randomly translate these patterns in the transverse plane to generate an augmented dataset (see \methods{}), with which we trained the different ANNs. Finally, we test our networks on unseen data, for which the target displayed by the DMD has been physically translated. In general, networks can retrieve the position $\theta$ of the target with precision that depends on many factors, including the optical thickness of the scattering media, the network architecture, and the number of training epochs. To quantitatively evaluate the precision of the ANN predictions, we compute the expected value of the target coordinates based on the distribution provided by the ANN (this is achieved using a softmax activation function, see \methods{}). Subsequently, we construct individual histograms for each target position and use them to calculate the corresponding standard deviations $\sigma_x$ and  $\sigma_y$, which we use as a figure of merit for the ANN precision. In addition, we calculate the average deviations of the predicted coordinates from the true positions to characterize the ANN biases $B_x $ and $B_y$.

\begin{figure*}[t]
    \begin{center}
		\includegraphics[width=0.9\textwidth]{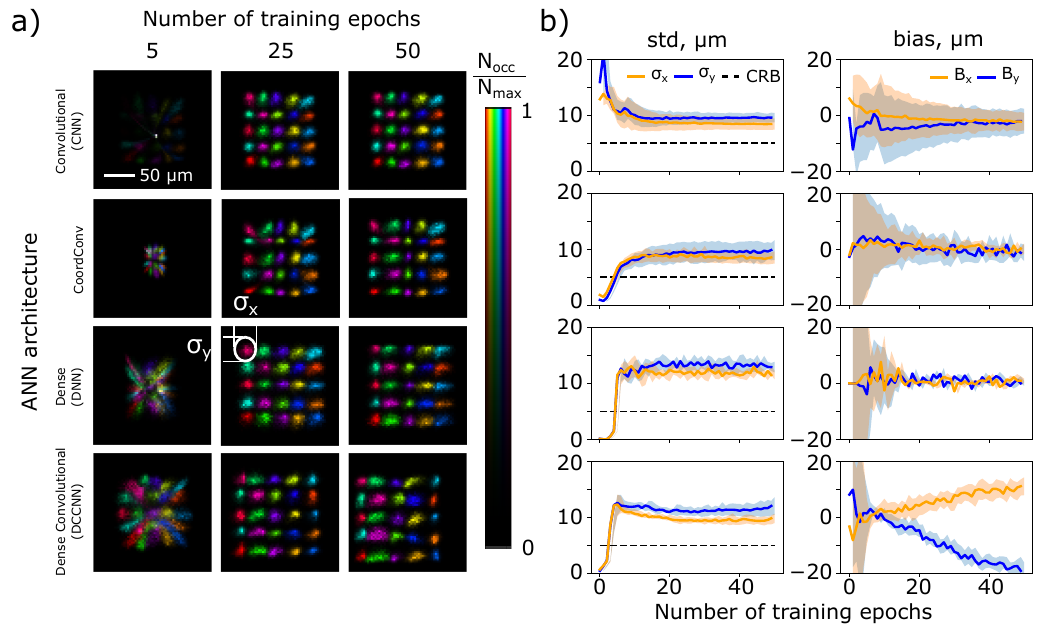}
	\end{center}
    \caption{Comparison of different ANN architectures for the sample with optical thickness $b = 4.2$. The rows in (a) and (b) correspond to respectively: Convolutional, CoordConv, Dense and Dense Convolutional Network architectures. (a) Two-dimensional histograms of the estimated position of the target. Different colors encode different ground truth positions, and brightness encodes probability. $N_\mathrm{max}$ = 5000 testing examples are used for each target position. The spread of the histograms characterizes the ANN precision. The columns show the progression of the prediction performance throughout the training process (after 5, 25 and 50 epochs respectively). (b) Evolution of the standard deviations (left column) and biases (right column) during training. The thick lines show the values averaged over all target positions, and the shaded areas represent the interquartile range (third minus first quartile) calculated over 25 positions. The black dashed lines in the left column represent the Cramér-Rao bound, calculated assuming that the estimators are unbiased.} 
	\label{fig_comparison}
\end{figure*}

The most straightforward ANN choice for image processing tasks is a convolutional neural network (CNN) \cite{goodfellow_deep_2016}. It has been shown, however, that ANNs with purely convolutional layers typically fail in the task of accurately tracking the coordinates of an object in an image~\cite{liu2018intriguing}. This is due to the intrinsic translational invariance of the convolution operation, which leads to the loss of information about the feature position. In order to break the spatial invariance of the convolution operation, one possibility is to explicitly add coordinate meshes to the layers (CoordConv layers), which leads to faster convergence and lower bias as compared to usual CNNs \cite{liu2018intriguing}. While fully connected layers (DNNs) are usually harder to train, they can also be used in our case since the target of interest has a simple spatial structure. Finally, a CNN modification featuring structured skip connections was recently introduced~\cite{8099726}. Such connections form a structure akin to a dense layer, leading to the name ``Densely Connected Convolutional Networks'' (DCCNNs), also known as ``DenseNets''.

We compared the performance of these four classes of ANNs. Prior to training, we performed hyperparameter tuning for each architecture (see \SI{4.2}). The best-performing ANN of each class was then trained and tested to locate the hidden target. The comparison of their performance is shown in \fig{fig_comparison}. In \ffig{fig_comparison}{a}, we present the estimated target positions for the four different architectures along the training process, for the scattering sample with optical thickness $b=4.2$. On these two-dimensional histograms, distinct ground truth target positions are represented by different colors, and the brightness of each color indicates the corresponding probability predicted by the ANN for that particular position. We observe that the choice of the ANN architecture does not seem to have a large influence over the estimated target position. To quantitatively analyze the performance of the different ANN architectures, we show in \ffig{fig_comparison}{b} the evolution of the standard deviations $\sigma_x$ and $\sigma_y$ (left column) and biases $B_x $ and $B_y$ (right column), calculated from the histograms in \ffig{fig_comparison}{a}. These values were obtained by averaging over statistical fluctuations of the ANN predictions. 

\begin{figure*}[t]
    \begin{center}
	\includegraphics[width=0.7\textwidth]{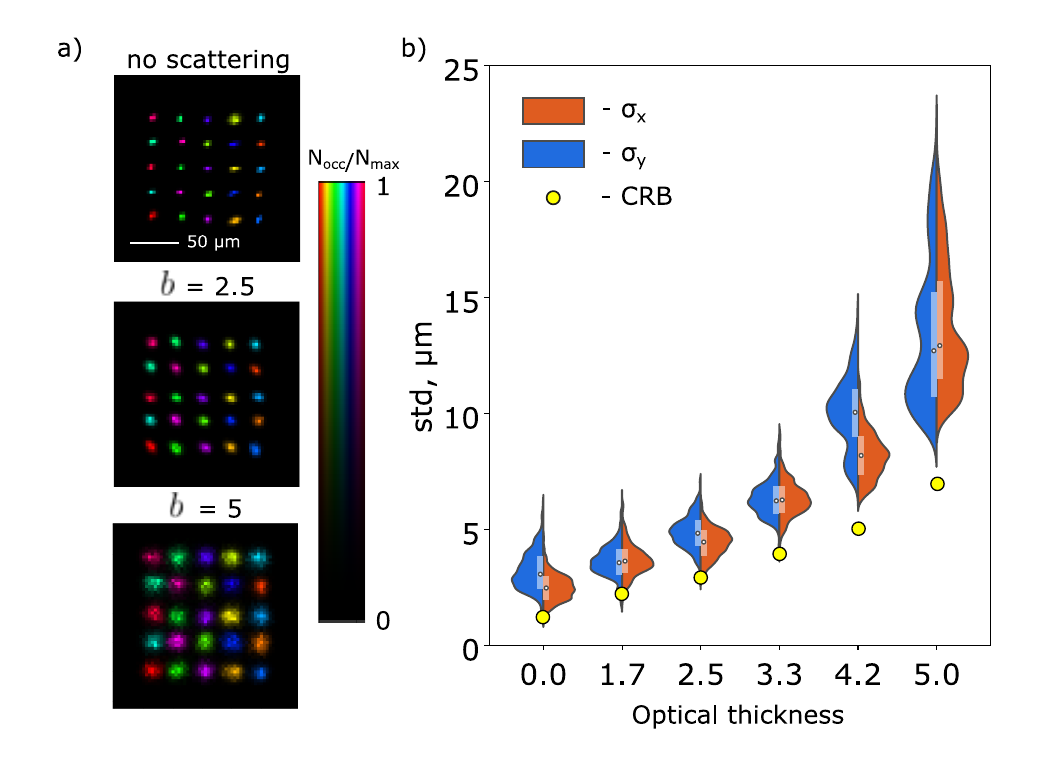}\end{center}
    \caption{Dependence of the ANN precision on the optical thickness of the scattering sample. (a) Examples of reconstructed histograms of the target positions using CoordConv architecture in the absence of scattering sample ($b=0$), as well as for $b = 2.5$ and $b = 5$. (b) Violin plots showing the CoordConv precision achieved for 25 random initializations of the network. White dots show the median values of $\sigma_x$ and $\sigma_y$, vertical bars indicate the corresponding first and third quartiles, and the colored areas show the associated histograms. The yellow circles represent the Cramér-Rao bound (CRB).}\label{fig_OD}
\end{figure*}

As it turns out, the CNN approaches the (unbiased) Cram\'er-Rao bound, which here is equal to 5.0\,\textmu m for the scattering sample under consideration ($b=4.2$). However, as can be seen from the large variability in the associated bias plot [\ffig{fig_comparison}{b}, first row, right column], the CNN develops a significant bias. This issue can be resolved by including coordinate layers, which strongly mitigates the bias while leading to the same standard deviation as the usual CNN. In the case of the DNN, the observed bias is smaller, but the standard deviation is larger than those of the CoordConv. The DCCNN architecture exhibit a very different behavior: after only a few epochs, it gives low bias and approaches the Cram\'er-Rao bound. Nevertheless, this advantage comes with a trade-off: a large bias is observed later in the training process, requiring careful consideration to stop the training at an appropriate point. Note that, in the case of the CoordConv, DNN and the DCCNN, the Cramér-Rao bound is overpassed in the first few epochs; this occurs because the initial guess of these architectures is zero for each coordinate, resulting in standard deviations equal to zero. Estimations are then highly biased in the early stage of the training, which makes the unbiased Cramér-Rao bound calculated using \eq{eq_crlb} not relevant to the estimations performed during these first epochs.

\bigskip


\subsection*{Achievable precision for different scattering strengths}

Based on the combination of minimal uncertainty and small bias (lowest mean squared error, see \SI{4.3}), we select CoordConv as our best architecture and study how it performs at different scattering strengths. As an illustration, we show in \ffig{fig_OD}{a} the ANN predictions in the absence of a scattering sample as well as for two other optical thicknesses ($b=2.5$ and $b=5$). While the bias seems unaffected by the presence of the scattering sample, the variance of the estimated positions greatly increases for thick scattering samples. To quantitatively describe this behavior, we show in \ffig{fig_OD}{b} the standard deviations achieved with the CoordConv architecture as a function of the optical thickness, obtained by training and testing this architecture 25 times with different random initial weights. These results also include a bias correction step, which is necessary in order to compare the ANN predictions to the unbiased Cramér-Rao bound (see \SI{4.4}).

As expected, the average standard deviation increases with the optical thickness, ranging from approximately 2.5\,\textmu m in the absence of a scattering sample up to 12.8\,\textmu m for a strongly scattering sample ($b=5$). Remarkably, for small optical thicknesses, the standard deviation of the ANN comes very close to the Cram\'er-Rao bound [yellow circles in \ffig{fig_OD}{b}], demonstrating that the CoordConv architecture can reach the ultimate precision limit. In some cases, it even seems that the Cramér-Rao bound can even be overpassed; this artifact that can be explained by the approximations made during the calculation of the Fisher information (see \SI{3.5}) and by the finite size of the test dataset that was used to calculate the standard deviation of the ANN (see \SI{4.5}). For larger optical thicknesses, we observe a large variability in the ANN precision, as being captured by broader histograms in \ffig{fig_OD}{b}. This variability comes essentially from the dependence of the ANN precision on the target position. We indeed observe that off-central positions, which are less connected to the output of the network, are harder to predict than central ones (see \SI{4.6}). Moreover, it is also plausible that, with a larger number of trainable weights, the ANN would perform better at high optical thicknesses (see \SI{4.2}).

In \ffig{fig_OD}{b}, we can also observe that the dependence of the Cram\'er-Rao bound on the optical thickness is approximately linear between $b=0$ and $b=5$. While no analytical expression is currently available to describe this behavior, we suppose that it arises here as the result of a competition between shot noise and mechanical vibrations of the sample (mostly relevant for small optical thicknesses), as well as exponential attenuation of ballistic photons (mostly relevant for large optical thicknesses).

Finally, we also performed cross-tests of the models, that we trained and tested using datasets associated with different optical thicknesses and different object sizes. We observe that models trained on the strongest scattering dataset retain their ability to predict the target's position in weakly scattering conditions, while the models trained on weak scattering dataset become imprecise to estimate the target position in strongly scattering conditions (see \SI{5.1}). Moreover, the models seem to generalize relatively well for small variations in the object size and shape (see \SI{5.2}).


\bigskip
\section*{Discussion}

We have shown that the Cram\'er-Rao inequality, which is the fundamental inequality that limits the precision of any estimator, also provides the ultimate benchmark for different ANN architectures trained to provide estimates on complex photonic systems. For the problem of estimating the position of a target behind a dynamic scattering sample, the performance of all architectures is similar, with a slight advantage for the convolutional neural network with CoordConv layers. We have then demonstrated that, for different optical thicknesses, this architecture approaches the Cramér-Rao bound.

Calculating the Cram\'er-Rao bound entails accessing the Fisher information matrix describing measured data. While this matrix is typically derived from analytical models, we presented here a general model-free approach to approximate the Fisher information matrix of an unknown statistical distribution based on experimental data. Note that, in general, the Fisher information matrix depends on the true value of the parameters $\theta$, and therefore the Cram\'er-Rao bound allows one to assess the local performance of ANNs around a given value in parameter space. While we studied here a system with translational invariance for which the Fisher information does not depend on $\theta$, our work could also serve as a building block to study more involved situations requiring the Fisher information to be calculated over the whole parameter space. Another perspective of our work is to analyze the achievable precision when large images need to be reconstructed. While a pixel-based imaging strategy seems unpractical here due to the large number of parameters that is typically involved \cite{barrett_objective_1995,bouchet_fundamental_2021}, finding relevant sparse representation of these images appears as a promising strategy \cite{szameit_sparsity-based_2012,bouchet_optimizing_2021}. Our analysis performed with two parameters (the transverse coordinates of the target) thus paves the way towards the analysis of more complex imaging scenarios in which many parameters need to be estimated.

The formalism inherently includes all physical effects that affect the precision with which parameter values can be estimated. In our experiments, the statistics of the noise were of fundamentally different origins. Indeed, in free space, the precision was essentially limited by shot noise as well as by mechanical vibrations of the experimental setup. In contrast, in the presence of complex scattering media, the precision was limited by the occurrence of random scattering events in the media. Yet, we have shown that the Cram\'er-Rao bound can be successfully calculated regardless of the physical origin of the noise, and without any analysis of the speckle correlations characterizing our scattering systems \cite{PhysRevLett.61.834,akkermans_mesoscopic_2007,starshynov_non-gaussian_2018}. In fact, the key quantity is rather the likelihood function, which is assessed directly from measured experimental data and without any assumption on the data statistics. Moreover, all prior information is naturally included in the parameterization of the system. As such, also far-field super-resolution effects are inherently described by the formalism. 

The method itself is very general and can be used to assess whether any artificial neural network is optimal with respect to the task it was designed for. We expect our approach to be especially useful to drive algorithmic developments in the field of computational imaging~\cite{barbastathis_use_2019}. Major potential applications include imaging through multimode fibers \cite{cao_controlling_2023} and through complex scattering tissues \cite{yoon_deep_2020}, for instance for in-vivo neuronal imaging \cite{stiburek_110_2023,sarafraz_speckle-enabled_2024}.


\section*{Methods}

\subsection*{Experimental setup}

The setup used in our experiments is shown in \SI{1}. A laser beam from a diode (Thorlabs DJ532-40) is expanded to a diameter of 2\,mm before entering the optical system. Lens L2 forms the image of the front surface of the scattering sample on a sCMOS camera (Andor Zyla 5.5). Lens L1 is placed in such a way that the input beam after L2 is collimated. Using the lens L3 and the microscope objective MO (Olympus Plan N 10x), we image the DMD (Vialux V-7001, pixel size 13.7\,\textmu m) onto the back surface of the scattering sample, while also forming the illumination area of 8\,mm diameter on a DMD. The resulting magnification of the optical system formed by L2, MO and L3 is 2.9. The DMD is oriented such that the surface of ON pixels is perpendicular to the optical axis of the imaging system, i.e., these pixels back-reflect the incident light towards the sample. In contrast, the surface of OFF pixels is tilted such that light reflected by these pixels escapes the optical system, because it goes outside the aperture of lens L3. Note that, in this configuration, the DMD is tilted by a few degrees in the $y$ direction, which leads to a slight defocus for different $y$ positions. 

As a scattering sample, we use a suspension of TiO2 nanoparticles (Sigma-Aldrich) in glycerol, which is pumped through the flow cuvette F (Helma, 6.2\,\textmu L, optical path length 100\,\textmu m) using the pressure chamber and compressor at a rate of around 7\,ml/hour. With this setup, the measured decorrelation time is around 30\,ms (see \SI{1}). We chose the time interval between two successive frames (33.3\,ms) so that two successive frames are uncorrelated, and we chose the exposure time (200\,\textmu s) so that the speckle is stable during the measurement of one frame. We use a DMD area of 5$\times$5 pixels (68.5\,\textmu m$\times$68.5\,\textmu m) as a model dynamic target and we record a sequence of reflected speckle patterns with different target positions, while the scattering liquid provides different realizations of the optical disorder for each speckle. With our optical setup, the DMD is directly imaged onto the sample plane, with a magnification factor of 0.18.

\subsection*{Data collection}

For each optical thickness, a total of $5\times10^5$ images composed of 128$\times$128 pixels (1 pixel corresponding to 6.5 \textmu m$^2$) were taken in the following sequence: B, S, B, D, B, S, B, D, \dots, where B represents the background speckle measured with all DMD pixels at 0 (i.e., none of the measured light comes from the DMD), S represents the static target dataset (with the target fixed at the center of the field of view), and D represents the dynamic target dataset (with the target moving in a snake pattern across a $5\times5$ grid of positions separated by 2 DMD pixels vertically and horizontally). 

The collected background images were used to subtract the slowly changing speckle pattern occurring due to TiO$_2$ particles attaching to the walls of the cuvette.  The time series of each pixel of the background dataset was filtered with a Savitzky–Golay filter (5000-sample window, or approximately 170\,s) to form a dynamic background signal, that is then subtracted from the consecutive static and dynamic dataset images. Note that the subtraction of the slowly-varying average background occurs on a time scale (170\,s) that is much larger than the decorrelation time (30\,ms), which means that background subtraction cannot allow us to isolate the ballistic light. We observed that, without background subtraction, ANNs perform much worse: indeed, since we use a data-augmentation procedure, ANNs incorrectly learn to rely on the numerically shifted background features during training. 

The static dataset images were numerically shifted horizontally and vertically by a random number of pixels (extracted from a uniform distribution) within the interval [-40,40] pixels. Both static and dynamic datasets were then resized to obtain 32$\times$32 images by pixel binning and were then normalized from 0 to 1 by dividing each image by its maximum intensity value. As a result of the procedure above, we obtained a training dataset of $1.25\times10^5$ images (32$\times$32 pixels) and a testing dataset of the same size with 25 target positions. Note that the size of the testing dataset has been chosen sufficiently large, in order for us to assess the variance of the ANN predictions ($5\,000$ patterns for each of the $25$ different positions of the target).

\subsection*{ANN structure and optimization}
We have observed that, for the task of predicting target coordinates on a discrete grid, a classification approach works better than a regression approach. For each architecture, the target's vertical and horizontal coordinates were one-hot encoded into vectors of $N_h$ length, which then were merged into a $2N_h$ length vector that served as the ground truth for the network. We employed categorical cross-entropy loss and soft-max activation for the final layer, enabling the network to produce the probability density of the target position. The final prediction for the position of the target was taken as the expected value of the horizontal and vertical position distributions computed from a single frame in the test dataset. For each architecture, we performed a hyperparameter grid search. The optimization space for each model was selected in such a way that the maximal number of parameters would be similar for different network architectures. The details about the search parameter space and the optimal hyperparameter values used in the training can be found in \SI{4.2}. 

The CoordConv architecture is found to be the optimal architecture for our task. Although the bias it develops is small, it is still non-negligible. We correct this bias by splitting a part of a training set, making predictions from it, and comparing it to the ground truth. We fit a 2D spline function to the set of points for which we can infer the bias in that way so that this function covers the whole field of view (see \SI{4.4}).

\subsection*{Numerical simulations}

To generate the Gaussian dataset, we sample 125\,000 data points from a multi-dimensional Gaussian distribution with independent and identically distributed entries, a number of dimensions $N_\text{dim}=50$, a mean value $\mu = \theta$ and a standard deviation $\sigma=1$. To introduce correlations, we apply a matrix consisting of ones along the main diagonal and the first off-diagonals to every data point. For image data, this would correspond to neighboring pixels being strongly correlated but distant pixels being independent. We then estimate the Fisher information of this correlated dataset for the parameter $\theta$, which is known analytically to be $\mathcal{J}(\theta)=N_\text{dim}$~\cite{van_trees_detection_2013}. To generate the non-Gaussian dataset, we repeat the procedure above with the difference that we apply an additional non-linear transformation $f(x)=x^3$ to each component of the correlated random variable. We choose this transformation as it preserves the Fisher information (this can be demonstrated, e.g., using singular value decomposition). Hence, the value of the Fisher information is still $\mathcal{J}(\theta)=N_\text{dim}$, and we can verify whether our method returns an estimated Fisher information that is close to the true Fisher information.

In the inset of \ffig{fig_fi_method}{b} and \ffig{fig_fi_method}{c}, we show the Gaussian and the non-Gaussian distributions, respectively, by showing point clouds of two generic components $X_1$ and $X_2$ of the 50 dimensional random variables $X_k$. While the Gaussian distribution has an elliptical structure due to the correlations, the non-Gaussian distribution clearly has a more complex shape. The blue curves in \ffig{fig_fi_method}{b} and \ffig{fig_fi_method}{c} correspond to the dependence of the estimator of the total Fisher information $\sum_k \widehat{\mathcal{J}}^k$ as a function of the step size $\Delta\theta$ of the finite difference scheme, for the Gaussian data and the non-Gaussian data, respectively. These values are normalized by the true value of the Fisher information (known analytically), such that a value of $1$ corresponds to a perfectly accurate estimate. 

In these simulations, the optimal value of the step size $\Delta \theta_\text{thr}$ is selected by choosing the value where the fluctuations of the second derivative of the curve become greater than those for large step sizes by a factor of $10$ (see \SI{3.4}). We observe that in both cases the estimated Fisher information evaluated at this step size is close to the true Fisher information.



%


\subsection*{Acknowledgements}

I.S.\ and D.F.\ acknowledge financial support from the UK Engineering and Physical Sciences Research Council (EPSRC grants No.\ EP/T00097X/1, EP/Y029097/1). D.F.\ acknowledges support from the UK Royal Academy of Engineering Chairs in Emerging Technologies Scheme. L.M.R.\ and S.R. were supported by the Austrian Science Fund (FWF) through Project No.\ P32300 (WAVELAND).

\subsection*{Data availability}

The data generated in this study and scripts for data processing have been deposited in the University of Glasgow’s repository for research data with the following URL: \href{https://doi.org/10.5525/gla.researchdata.1926}{https://doi.org/10.5525/gla.researchdata.1926}.


\onecolumngrid
\pagebreak
\beginsupplement
\begin{center}
	\textbf{\large Model-free estimation of the Cramér-Rao bound \\ for deep-learning microscopy in complex media\\ \bigskip (Supplementary information)}
	
	\bigskip
	Ilya Starshynov,$^1$ Maximilian Weimar,$^2$ Lukas M.\ Rachbauer,$^2$\\ Günther Hackl,$^2$ Daniele Faccio,$^1$ Stefan Rotter,$^2$ and Dorian Bouchet$^3$\\ \vspace{0.15cm}
	\textit{\small $^\mathit{1}$School of Physics and Astronomy, University of Glasgow, Glasgow, G12 8QQ, United Kingdom}\\
	\textit{\small $^\mathit{2}$Institute for Theoretical Physics, Vienna University of Technology (TU Wien), 1040 Vienna, Austria}\\
	\textit{\small $^\mathit{3}$Univ. Grenoble Alpes, CNRS, LIPhy, 38000 Grenoble, France}
\end{center}
\vspace{0.3cm}

\section{Optical setup}
\label{sec_setup}

\begin{figure*}[h]
	\begin{center}
		\includegraphics[width=0.75\textwidth]{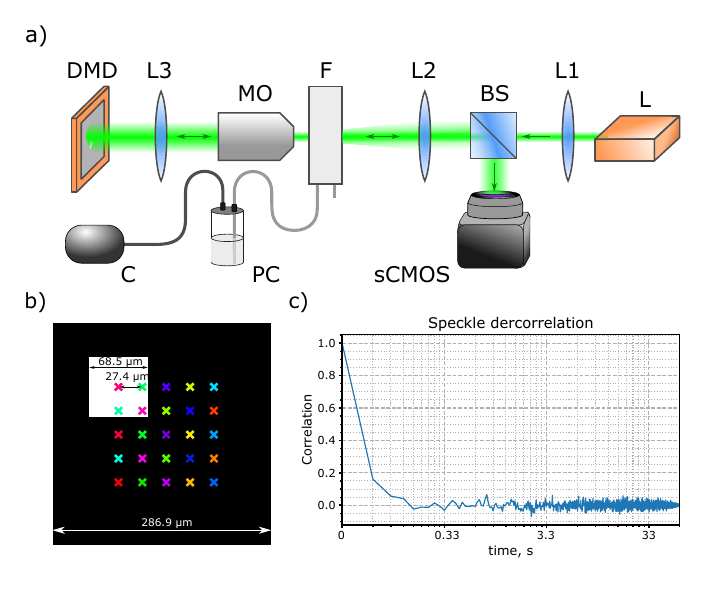}
	\end{center}
	\caption{(a)~Experimental setup: L - laser diode (Thorlabs DJ532-40), L1, L2, L3 - lenses (focal length $f = 100$\,mm), BS - 50/50 beam splitter, F - flow cuvette (Helma, 6.2\,\textmu L, optical path length 100\,\textmu m), DMD - digital micromirror device (Vialux V-7001), MO - microscope objective (Olympus Plan N 10x), PC - pressure chamber, C - compressor, sCMOS - camera (Andor Zyla 5.5). (b)~Sketch representing the dimensions of the target (in white), of the field of view (in black) and of the displacements used for the test set (colored crosses). The target is composed of 5$\times$5 DMD pixels and is displaced by 2 pixels between two adjacent positions. Dimensions are given here in the DMD plane (a magnification factor of $\times 0.18$ should be applied to calculate corresponding dimensions in the sample plane). (c)~Normalized intensity correlation function of the speckles measured in the absence of the target. The observed decorrelation time is around 30\,ms, which is of the order of the time interval between successive frames (33.3\,ms) and much longer than the exposure time of the camera (200\,\textmu s).}
	\label{fig_setup}
\end{figure*}

	\section{Estimating the target position from ballistic and scattered light}
	\label{sec_ballistic}
	
\begin{figure*}[!ht]
	\begin{center}
		\includegraphics[width=0.9\textwidth]{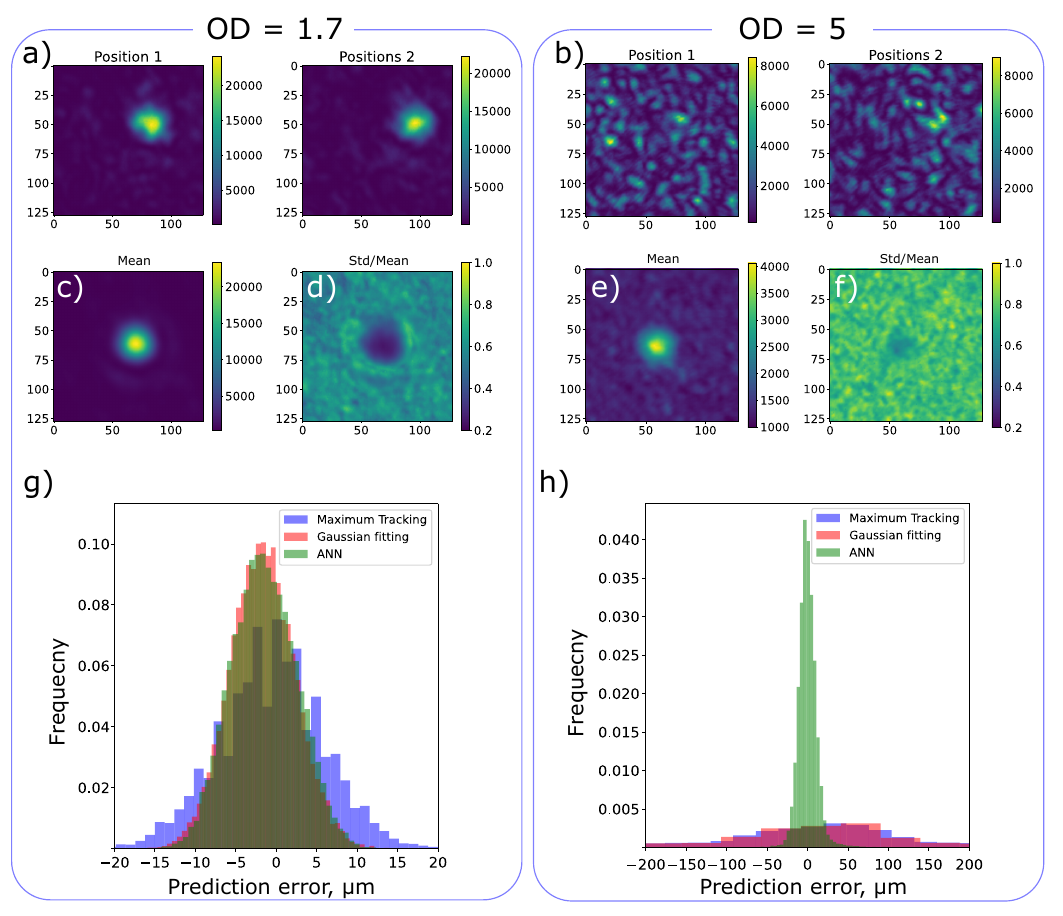}
	\end{center}
	\caption{(a),(b) Single shot intensity images extracted from the test set for two adjacent positions of the target, for optical thickness of (a) $b =1.7$ and (b) $b =5$. In the stronger scattering case, brighter spots are typically observed around the target position, but precisely localizing the target from such images is a difficult task. (c),(e) Average intensity calculated over 1000 frames, with the target located at the central position. (d),(f) Ratio between the standard deviation and the mean, which is equal to 1 for fully-developed speckles. A strong ballistic contribution can be seen in (d), as the ratio drops to 0.2. However, in (f), the minimal value is 0.85, which indicates that the contribution of the ballistic light is small as compared to the scattered one. (g),(h) Distribution of errors for different methods: maximum tracking, Gaussian fitting and ANN (CoordConv). In the strongly scattering case, the ANN performs much better than any of the simpler methods. In all images [(a) to (f)], axes are in camera pixels (6.5 \textmu m).}
	\label{fig:SI2_diffuse}
\end{figure*}

	For small optical thicknesses, we essentially measure the ballistic light coming from the target: the target position is visible on single-shot intensity images [\ffig{fig:SI2_diffuse}{a}], and the average intensity collected at the target position is significantly larger than the random fluctuations caused by the scattered light [\ffig{fig:SI2_diffuse}{c} and \ffig{fig:SI2_diffuse}{d}]. However, for larger optical thicknesses, the target position cannot be easily estimated from single-shot intensity images [\ffig{fig:SI2_diffuse}{b}] due to the presence of scattered light. A ballistic contribution remains, as can be seen on the average image [\ffig{fig:SI2_diffuse}{e}], but this contribution is significantly smaller than the random fluctuations caused by the scattered light  [\ffig{fig:SI2_diffuse}{f}]. In these conditions, the problem of detecting the exact position of the object becomes quite complicated, even in the presence of ballistic light. To illustrate this complexity, we have tested simple approaches based on maximal tracking and Gaussian fitting [\ffig{fig:SI2_diffuse}{g} and \ffig{fig:SI2_diffuse}{h}]; both of them fail at precisely estimating the target position at large optical thickness, while artificial neural networks are able to reach a much higher precision.

\section{Estimation of the Fisher information}
\label{sec_fi_calculation}

	\subsection{Elimination of statistical dependence in-between pixels}
	\label{sec_statistical_dependence}
	
	Due to the finite extent of speckle grains, values measured by neighboring pixels are correlated to each other. While Eq.\ (2) of the manuscript intrinsically includes the effect of these correlations, it is in practice much more robust to first remove the dependencies by performing an appropriate change of basis. 
	
	Among those linear transformations that meet the task of decorrelating the random variable, we identify two particularly practical types of transformations. First, similar to the technique of principal component analysis \cite{PCA_2}, we can estimate the covariance matrix of the random variable from the dataset and then transform the basis such that the covariance matrix becomes diagonal. By doing so, the lowest order dependencies (i.e. those captured by the covariance matrix) between different components of the transformed random variable are minimized. However, there could be higher order dependencies between these components, which would not be captured by the covariance matrix. While diagonalizing the covariance matrix is sufficient in many applications, it is possible to take these higher order dependencies into account even when restricting our analysis to linear transformations. The independent component analysis (ICA) \cite{hyvarinen2000_2} allows us to construct a linear transformation that minimizes the dependencies between components of the random variable. Unlike the technique used in principal component analysis, ICA harnesses more degrees of freedom when choosing the transformation matrix since there are fewer restrictions in the choice of the matrix (such as those related to the symmetry of the matrix).
	
	In general, when dealing with data from an unknown distribution, one needs to first test whether the data follow Gaussian statistics. Whenever the data are Gaussian, ICA is not applicable. In such a case, however, the simpler alternative to the ICA, that is estimating the covariance matrix and decorrelating the random variable by rotating the basis such that the covariance matrix is diagonal, is sufficient to guarantee statistical independence between the transformed random variables. For non-Gaussian distributions, dependencies between components cannot be completely removed even by diagonalizing the covariance matrix; the ICA must be chosen in such cases.
	
	In our experiments, we observe that the distributions are clearly non-Gaussian, hence we choose the ICA as a means of decorrelation. We thus apply an ICA in order to find the transition matrix $A$ transforming a random vector $X$ into its counterpart $Y=AX$ with approximately independent components. ICA requires that most of the components of $Y$ do not follow a Gaussian distribution, and a suitable measure of non-Gaussianity is maximized in the process. We employ a computationally efficient version of ICA, the FastICA algorithm as described in \cite{hyvarinen2000_2}. Here, after whitening the random vector's components to unit variance, non-Gaussianity is maximized iteratively until convergence is reached. As a result of ICA, the probability density function of the transformed random vector $Y$ approximately factorizes:
	\begin{equation}
		p(Y;\theta) \simeq \prod_{k=1}^N p_k(Y_k;\theta)\;,
		\label{eq_factorization}
	\end{equation}
	which allows us to employ the following simplified expression for the Fisher information matrix:
	\begin{equation}
		\left[\mathcal{J}(\theta)\right]_{ij} \simeq \sum_{k=1}^N \left \langle   \left[ \frac{\partial \ln p_k(Y_k;\theta)}{\partial \theta_i} \right] \left[ \frac{\partial \ln p_k(Y_k;\theta)}{\partial \theta_j} \right] \right \rangle \; ,
		\label{eq_independent}
	\end{equation}
	where $p_k(Y_k;\theta)$ are the marginal distributions of the individual independent components.

\subsection{Mutual information analysis of the independent components}

\label{sec_mutual_info}

Performing an independent component analysis (ICA) does not guarantee that the components $Y_i$ of the transformed random variable $Y=AX$ are independently distributed. Indeed, any given distribution is not necessarily related to a distribution with independent components by a linear transformation. In fact, since the speckle patterns follow complicated and unknown statistics, we do not expect the distribution $p(Y|\theta)$ to factorize exactly but only approximately. Modeling the distribution by a product requires us to employ a measure of dependence between components of $Y$. This is achieved by the concept of mutual information (MI) \cite{cover1999elements_2} between pairs of components of $Y$, which is defined as follows:
\begin{equation}
	I(Y_i,Y_j) = \int dX \int dY p(Y_i,Y_j) \log\left( \frac{p(Y_i,Y_j)}{p(Y_i)p(Y_j)}\right)\;,
	\label{eqn:MutualInformation}
\end{equation}
where $p(Y_i,Y_j)$ is the joint distribution of the components $Y_i$ and $Y_j$ and $p(Y_i)$ and $p(Y_j)$ are their marginal distributions, respectively.
Ideally, the MI between all distinct components vanishes since this corresponds to independently distributed components. We relax this condition by requiring that the ratio
\begin{equation}
	g_{i,j} = \frac{ I(Y_i,Y_j)}{I(Y_i,Y_i)}\;
	\label{eqn:MICondition}
\end{equation}
is small for all pairs of components $(Y_i,Y_j)$, where the denominator coincides with the entropy of the random variable $Y_i$. $g_{i,j}$ measures the MI of a pair of pixels, relative to the entropy of one of the pixels. Moreover, we expect that the sum of this quantity over all components $\sum_j g_{i,j}$ is not too large. The MI is estimated using histograms (1-dimensional histograms for the marginal distributions and 2-dimensional histograms for the joint distributions). For each fit, we use 10 bins of equal size.
For each data set we use $50 \ 000$ data points to construct the linear transformation by the ICA and reserve $75 \ 000$ for the estimation of the MI. 
The worst case result of the sum $\sum_j g_{i,j}$, i.e., the largest value among all components $Y_i$ is $\mathrm{max}(\sum_j g_{i,j}) \simeq (35.3, 3.2, 3.0, 2.6, 1.0, 0.5)$ for the Fisher information estimates shown in Fig.~3 of the manuscript, with the optical thicknesses $b=(0, 1.7, 2.5, 3.3, 4.2, 5)$ in this order. We observe that in the case of $b=0$, some components of the random variable contain a considerable amount of information about the other components. Here, modeling the data with a product of marginal distributions should be considered as a rough approximation.

\subsection{Derivation of the Fisher information estimator of the marginal distributions}
	
	\label{sec_fi_marginal}
	
	Applying \eq{eq_independent} entails estimating the probability density functions $p_k(Y_k;\theta)$ from a finite number of disorder configurations. This can be achieved by using histograms as approximate representations of $p_k(Y_k;\theta)$. In contrast to the conventional approach in density estimation where all bins are equal in width, using bins with equal frequency (i.e., equal fillings) yields a more stable estimate of the Fisher information, because small variations of the frequency do not influence the result.
	
	Then, in order to approximate the probability density functions $p_k(Y_k;\theta)$ for different values of $\theta$ close to $\theta=(0,0)$, we take advantage of the transverse spatial invariance of the problem instead of physically moving the target. This allows us to estimate the partial derivatives of $p_k(Y_k;\theta)$ with respect to the parameters of interest with a centered finite difference scheme
	\begin{equation}
		\frac{\partial p_k(Y_k;\theta)}{\partial \theta_i}
		\simeq
		\frac{ p_k(Y_k;\theta + \hat{e}_i\Delta \theta)- p_k(Y_k;\theta-\hat{e}_i \Delta \theta)}{2\Delta \theta}\;,
		\label{eqn:finite_difference}
	\end{equation}
	where $\hat{e}_i$ is the $i$-th unit vector in the parameter space (this space is of dimension $2$ in our case). To apply this procedure in practice, we first resize the images from $128\times 128$ pixels to $32\times 32$ pixels (which is the same resolution used for training the ANNs), and we then shift the images by $\Delta \theta$. Note that this approach results in a greater numerical stability as compared to resizing the images after the shift. We attribute this behavior to the fact that the resizing involves averaging over neighboring pixels; statistical fluctuations have a smaller influence on the resized pixels, which simplifies the interpolation between values of consecutive pixels that occurs when the images are shifted.
	
	Our next task is to construct an estimator for the Fisher information of each marginal distribution $p_k(Y_k;\theta)$ using the finite difference approximation, based on the statistical sample that is available. Using statistical samples of the $k$-th marginal distribution $p_k(Y_k;\theta)$ as well as of the distributions $p_k(Y_k;\theta\pm\hat{e}_i\Delta\theta)$, we can estimate the Fisher information by constructing histograms for the distribution $p_k(Y_k;\theta)$, choosing the size of the bins such that the filling of the $j$-th bin $h_j^k=h^k$ is the same for each bin $i$. We denote the fillings of $p_k(Y_k;\theta\pm\hat{e}_i\Delta\theta)$ using the notation $h_j^{\pm, k}$, respectively. We assume that all samples are of the same size, which implies that $\sum_jh_j^k=\sum_jh_j^{\pm, k}=H$. With the width $\Delta x_j$ of each bin that is determined by the filling $h_j^k$, we can approximate the probability density functions (PDFs) at the center of each bin by $p^k_j\simeq h_j/(H\Delta x_j)$ and $p^{\pm, k}_j\simeq h_j^{\pm, k}/(H\Delta x_j)$. The Fisher information with respect to the parameter $\theta$ can be written in a convenient way by
	\begin{equation}
		\mathcal{J}_{ii}^k(\theta) = \int dY_k p_k(Y_k; \theta) \left( \frac{\partial \ln p_k(Y_k; \theta)}{\partial \theta_i} \right )^2  = 4 \int dY_k \left( \frac{\partial \sqrt{ p_k(Y_k; \theta)}}{\partial \theta_i} \right )^2\;. 
		\label{fi_single}
	\end{equation}
	Plugging the discretized estimate of the PDF into this equation and approximating the derivative by a symmetric finite difference yields
	\begin{eqnarray}
		\mathcal{J}_{ii}^k(\theta)  \simeq 4 \sum_j \left( \frac{\Delta x_j}{2\Delta\theta}\left (\sqrt{p^{+, k}_j}-\sqrt{p^{+, k}_j}\right)\right)^2 = 
		4 \sum_j \left( \frac{\Delta x_j}{2\Delta\theta}\left (\sqrt{\frac{h_j^{+, k}}{\Delta x_j H}}-\sqrt{\frac{h_j^{-, k}}{\Delta x_j H}}\right)\right)^2\;.
	\end{eqnarray}
	Based on this expression, we can thus identify the following estimator for the Fisher information of the marginal distribution $p(Y_k;\theta)$:
	\begin{equation}
		\widehat{\mathcal{J}}_{ii}^k = \frac{1}{H(\Delta \theta)^2}\sum_j \left( \left (\sqrt{h_j^{+, k}}-\sqrt{h_j^{-, k}}\right)\right)^2 \;.
		\label{FI_estimator}
	\end{equation}
	The total Fisher information is estimated by summing all contributions from the marginals using $\widehat{\mathcal{J}}_{ii} = \sum_k\widehat{\mathcal{J}}_{ii}^k$. In \eq{FI_estimator}, we recall that the index $i$ denotes either the $x$ or the $y$ position of the target; this index enters the equations above only via the direction in which the images are shifted for the finite difference approximation. In the same way as for \eq{fi_single}, the off-diagonal elements of the Fisher information matrix are expressed by
	\begin{equation}
		\mathcal{J}_{ij}^k(\theta) = \int dY_k p_k(Y_k; \theta) \frac{\partial \ln p_k(Y_k; \theta)}{\partial \theta_i}\frac{\partial \ln p_k(Y_k; \theta)}{\partial \theta_j} \;,
	\end{equation}
	where $i\neq j$. The resulting estimator for these off-diagonal elements is
	\begin{equation}
		\widehat{\mathcal{J}}_{ij}^k = \frac{1}{H(\Delta \theta)^2}\sum_l \left (\sqrt{h_{il}^{+, k}}-\sqrt{h_{il}^{-, k}}\right) 
		\left (\sqrt{h_{jl}^{+, k}}-\sqrt{h_{jl}^{-, k}}\right)
		\;,
	\end{equation}
	where the indices $i$ and $j$ in $h_{il}^{\pm, k}$ and $h_{jl}^{\pm, k}$ now denoting that the images where shifted in the direction corresponding to $i$ and $j$, respectively. In practice, in our experiments, we observed that these off-diagonal elements are negligible compared to the diagonal elements, and we thus treat the problem as being effectively a single parameter estimation problem.

\subsection{Selection of the step size}
	
	\label{sec_step_size}
	
	Due to statistical fluctuations affecting histogram bin fillings, the approximation of the derivative and thus the estimate of the Fisher information depend on the value of the step size $\Delta \theta$. These fluctuations have a strong impact if the step size is too small. In contrast, if the step size is too large, we expect the finite difference approximation to be less accurate. Due to this trade-off, the optimal choice for $\Delta\theta$ is the smallest possible value for which the finite difference estimate of the Fisher information is not dominated by noise. We assume that this is the case when the estimate is stable with respect to small variations of $\Delta\theta$. To find this optimal value, we calculate the estimated Fisher information as a function of $\Delta\theta$ and observe that the estimated Fisher information is large for small $\Delta\theta$ and falls off rapidly when $\Delta\theta$ is increased. For larger $\Delta\theta$, the curves become almost independent on the step size and reach a plateau. We expect that the optimal step size is in the latter region since the finite difference approximation must not strongly depend on the choice of the step size. While one can easily find a reasonable step size by eye (by selecting one value where the curve flattens out), we employ here a formal criterion based on the second derivative of the curve, which allows us to identify the plateau as the region for which the slope remains constant. The second derivative fluctuates strongly for small $\Delta\theta$, while these fluctuations become small for larger $\Delta\theta$ when the plateau is reached. We then calculate the standard deviations of these fluctuations observed for large step sizes (when the plateau is already reached) by selecting a small window around a given step size. Our final selection of the optimal step size is the one where this standard deviation becomes greater than 10 times the standard deviation for the largest $\Delta\theta$ in the curves. In practice, this criterion corresponds to the selection of the smallest step size that belongs to the plateau.

\begin{figure*}[ht]
	\begin{center}
		\includegraphics[width=0.75\textwidth]{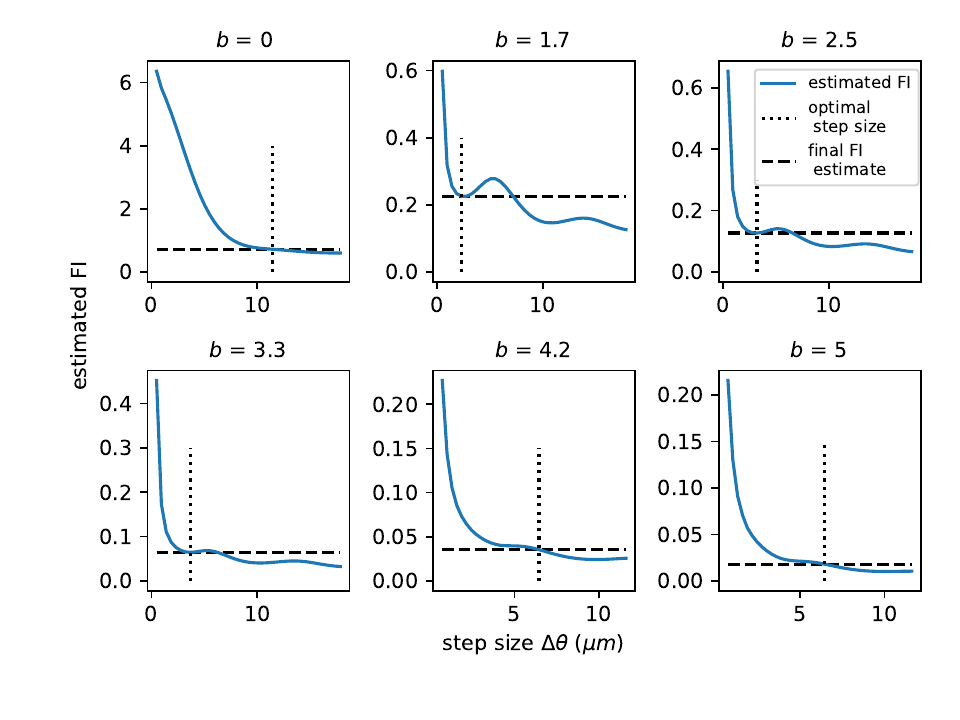}
	\end{center}
	\caption{The blue curves depict the estimated Fisher information for the experimental datasets as a function of the step size $\Delta \theta$ for different optical thicknesses $b$. The Fisher information is estimated using ICA for decorrelation and histograms to find the marginal densities. The vertical dotted lines correspond to step sizes where the Fisher information estimate shows good stability.}
	\label{fig_CRLB_vs_delta}
\end{figure*}

In \fig{fig_CRLB_vs_delta}, we show the estimated Fisher information for the experimental datasets with different optical thicknesses. We observe that the curves show some features that are more complex as compared to numerically generated data [Fig.~2(b) and Fig.~2(c) of the manuscript]. However, they follow the same trend: a rapid decrease of the estimated Fisher information for small values of $\Delta\theta$, and a slow decrease of the estimated Fisher information for larger $\Delta\theta$. For cases in which a (tilted) plateau can be observed, we use the smallest step size where the second derivative of the curve vanishes, which corresponds to the onset of the plateau. If the second derivative does not vanish anywhere (as observed for an optical thickness of $b=0$), we employ the same threshold as for the numerically generated data to identify the onset of the plateau. Finally, when oscillations are observed, we choose the location of the first local minimum to approximate the location of the onset of the plateau.

	\subsection{Uncertainty in the estimated Cramér-Rao bound}
	
	\label{sec_fi_uncertainty}
	
	Our method to estimate the Cramér-Rao bound from experimental measurements is inherently subject to several sources of errors. 
	
	A significant source of error comes from the approximation of the derivatives by a finite difference scheme and the associated selection of the optimal step size. In \fig{fig_CRLB_vs_delta}, we show the estimated Fisher information as a function of the step size in our experiments, for different optical thicknesses. From these curves, we can get an estimate of the error in the estimated Cramér-Rao bound by considering the maximum fluctuations of the curve in the region between the first and the second minimum (considering larger step sizes as being too large for the finite difference approximation to hold). In practice, we calculate the difference between the largest and the smallest Cramér-Rao bound in this region of step sizes, and we divide it by our estimated Cramér-Rao bound to obtain a relative error. The resulting error is typically around 30 \%, but with a value of around 10 \% for the optical thickness $b=0$ and a value of around 40 \% for the optical thickness $b=5$. Note that, while this uncertainty in the absolute value of the Cramér-Rao bound is significant, the uncertainty in the relative values of the bound calculated from similar datasets is much lower (of the order of a few percents).
	
	\begin{figure*}[ht]
		\begin{center}	\includegraphics[width=0.35\textwidth]{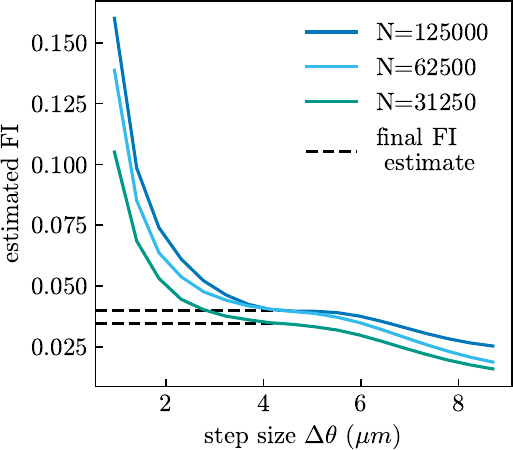}
		\end{center}
		\caption{The estimation of the Fisher information by selecting the optimal step size is shown for different sample sizes $N$. The curves depict the Fisher information as a function of the step size $\Delta \theta$ for one example dataset (optical thickness $b=4.2$). The horizontal dashed lines indicates the final estimates of the Fisher information where the curves show the first plateau.}
		\label{fig_CRLB_vs_delta_vs_sample}
	\end{figure*}
	
	Another possible source of error is due to the finite number of samples used to estimate the probability densities. However, the size of our datasets is sufficient so that we can neglect this source of error, both in simulations and in experiments. In simulations, we can verify this by repeating the whole analysis 100 times with different noise realizations, and calculate the mean $\mu_{\mathrm{FI}}$ and the standard deviation $\sigma_{\mathrm{FI}}$ of the estimated Fisher information. We find $\mu_{\mathrm{FI}}/\mathcal{J} = 1.01$ and $\sigma_{\mathrm{FI}}/\mathcal{J} = 0.05$ for the Gaussian distribution and $\mu_{\mathrm{FI}}/\mathcal{J} = 0.97$ and $\sigma_{\mathrm{FI}}/\mathcal{J} = 0.03$ for the non-Gaussian distribution. In the experiments, in order to check the influence of the size of the dataset, we used only a fraction of the original datasets and verify that the estimated Fisher information remains constant. We observed that our choice of the optimal step size and the corresponding Fisher information estimate does not change with smaller sample sizes, unless the sample size becomes very small. As an example, we show in \fig{fig_CRLB_vs_delta_vs_sample} the procedure of estimating the Fisher information by selecting the optimal step size for a single dataset ($b=4.2$) but for different sizes of the dataset. We observe that, taking either the full dataset (with a size of $125000$) or only half of it, we obtain nearly the same location for the first plateau of the curve curve, and thus nearly the same estimated Fisher information. When using only a fourth of the full dataset, the estimated Fisher information decreases slightly but still yields a reasonable value, the plateau being less pronounced in this case.

	Finally, note that another source of error could be due to the fact that the ICA does not perfectly decorrelate the data. However, this source of error is unlikely to significantly affect our estimates of the Fisher information. Indeed, we have observed in simulations that the true Fisher information is reached using our approach even for correlated data (see Fig.~2 of the manuscript). Moreover, in the experiments, we observed that the mutual information of the transformed variables is small (see Section \ref{sec_mutual_info}), which indicates that the ICA effectively managed to remove statistical dependence in-between the components of the transformed variables.

\section{ANN architecture}

\subsection{Data processing procedure}

\label{sec_data_processing}

\begin{figure*}[!ht]
	\begin{center}
		\includegraphics[width=\textwidth]{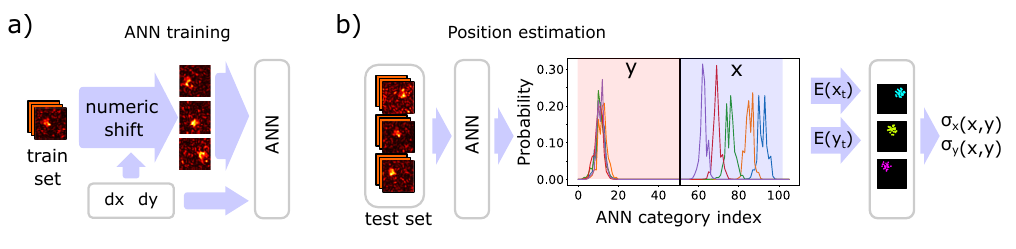}
	\end{center}
	\caption{(a) ANNs are trained using data augmentation by numerically shifting data measured with the target located at the central position. (b) ANNs predict the $x$ and $y$ coordinates of the target as a probability density, thanks to the softmax activation function used in the last layer. As an example, we show the probability densities given by the ANN for target positions in the top row ($b=4.2$ dataset). In this example, the $x$ coordinate of the target changes, while the $y$ coordinate remains fixed. The left 53 category indices correspond to the $y$ coordinate estimates, and the right 53 category indices correspond to the $x$ coordinate estimates. The actual target position is inferred as the expected values of those distributions. The achieved precision is calculated as the standard deviation of the statistics of the predicted position for different test examples.}
	\label{ANN_training_tes}
\end{figure*}

\subsection{Hyperparameter tuning}

\label{sec_hyperparameter}

\begin{figure*}[htb]
	\begin{center}
		\includegraphics[width=0.65\textwidth]{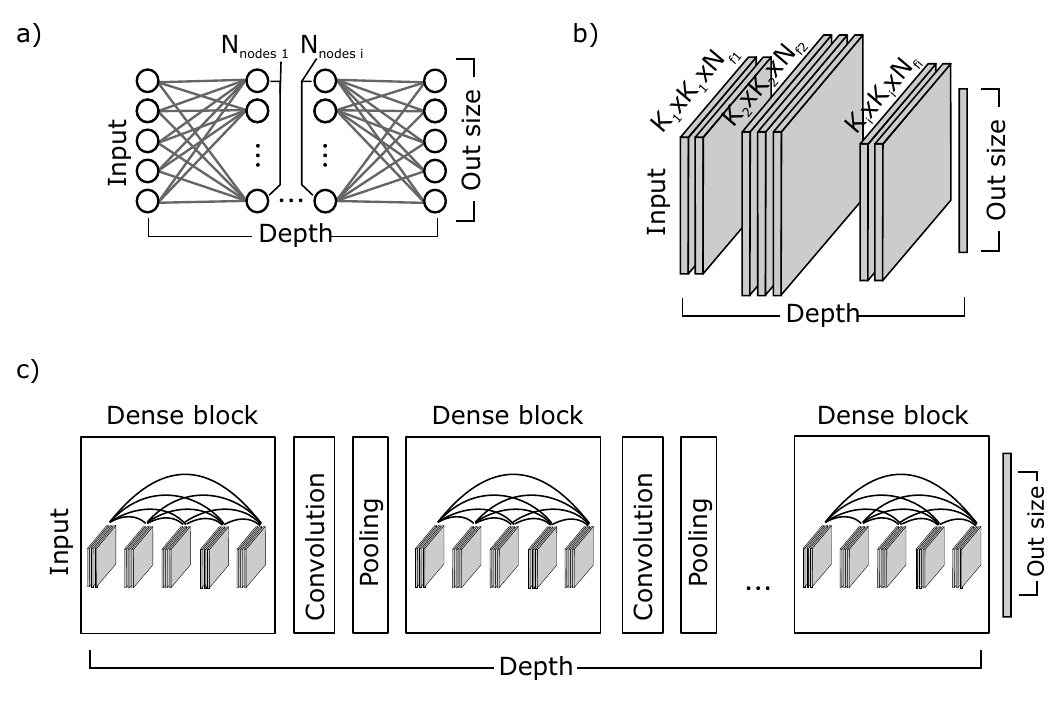}
	\end{center}
	\caption{ANN layouts. (a) Dense (DNN). (b) Convolutional (CNN), for the Convolutional with coordinate layers (CoordConv) the two coordinate layers (rescaled -1 to 1) were added after each layer stack. (c) Densely Connected Convolutional (DCCNN).   }
	\label{fig_Anns}
\end{figure*}

As the accuracy and the precision of the predictions may depend on the network architecture, we had to tune the ANN's hyperparameters to achieve optimal performance. The layouts of the architectures we tested (Dense, Convolutional, Convolutional with coordinate layers, and Dense Convolutional) are shown in \fig{fig_Anns}. The parameters to be tuned for each architecture and for $b=4.2$ are outlined in Tables~\ref{table_dnn} to \ref{table_DCCNN}. The dependence of the optimal CoordConv hyperparameters on the optical thickness is outlined in Table~\ref{table_OD}. 

For each of the networks, the ``Depth'' parameter denotes the number of layers, and the ``Out size'' parameter specifies the discretization of the predicted position probability distribution. The detailed description of the DCCNN hyperparameters can be found in Ref.~\cite{8099726_2}. The parameters were optimized using a grid search: for each combination, the corresponding ANN has been trained for 3 epochs and the values of loss, validation loss and prediction MSE were recorded. Using this strategy, we have observed that the CoordConv architecture performs best for a number of trainable parameters that is approximately constant over all optical thicknesses (around $5\times 10^5$ parameters, while the number of trainable parameters was swept from $1.6\times 10^5$ to $96.7\times 10^5$ for each optical thickness). However, the automatic hyperparameter tuning strategy is based on 3 epochs, which might not be representative of the behavior of the ANNs at the end of the training; it thus remains plausible that, with a larger number of parameters, the network could perform better at high optical thicknesses.

\begin{table}[!htb]
	\begin{center}
		\begin{tabular}{||c c c ||} 
			\hline
			Parameter & Tuning range & Optimal value  \\ [0.5ex] 
			\hline\hline
			Depth & 2-12 & 10 \\ 
			\hline
			Out size & 13-43 & 13 \\
			\hline
			Number of nodes in layers 1-2 & 20-820  & 420 \\
			\hline
			Number of nodes in layer 2-10 & 20-820  & 420 \\ [1ex] 
			\hline
			Trainable parameters $\times10^5$ & 4.2-86.9 & 8.7\\
			\hline
		\end{tabular}
		\caption{\label{table_dnn}Tuning ranges and optimal parameters of the DNN.}
	\end{center}
\end{table}

\begin{table}[!htb]
	\begin{center}
		\begin{tabular}{||c c c c ||} 
			\hline
			Parameter & Tuning range & Optimal value CNN & Optimal value CoordConv  \\ [0.5ex] 
			\hline\hline
			Depth & 2-5 & 4 & 2 \\ 
			\hline
			Out size & 13-63 & 53 & 58 \\
			\hline
			Filter shape  & 6$\times$6 - 10$\times$10 &  7$\times$7 &  8$\times$8 \\
			\hline
			Number of filters for each layer & 20-100  & 90 & 20 \\
			\hline
			Trainable parameters $\times10^5$ & 1.6-96.7 & 3.3 & 5.5\\
			\hline
		\end{tabular}
		\caption{\label{table_cnn}Tuning ranges and optimal parameters of the CNN and CoordConv}
	\end{center}
\end{table}

\begin{table}[!htb]
	\begin{center}
		\begin{tabular}{||c c c ||} 
			\hline
			Parameter & Tuning range & Optimal value  \\ [0.5ex] 
			\hline\hline
			Depth & 2-6 & 5 \\ 
			\hline
			Out size & 13-63 & 33 \\
			\hline
			Dense block depth & 2 - 6 &  2 \\
			\hline
			Dense block growth & 20-60  & 30 \\
			\hline
			Dense block bypass & 20-60 &  30 \\
			\hline
			Trainable parameters $\times10^5$ & 3.4-240 & 35\\
			\hline
			
		\end{tabular}
		\caption{\label{table_DCCNN}Tuning ranges and optimal parameters of the DCCNN.}
	\end{center}
\end{table}

\begin{table}[!htb]
	\begin{center}
		\begin{tabular}{||c c c c c c c||} 
			\hline
			Parameter & none & b=1.7 & b=2.5 & b=3.3 & b=4.2 & b=5  \\ [0.5ex] 
			\hline\hline
			Depth & 3 & 4  & 3  & 3  & 2  & 2 \\ 
			\hline
			Out size & 53 & 58 & 13 & 48 & 58 & 23 \\
			\hline
			Filter shape & 7$\times$7 & 6$\times$6 & 6$\times$6 & 6$\times$6 & 8$\times$8 & 7$\times$7 \\
			\hline
			Number of filters for each layer & 20 & 20 & 30 & 20 & 20 & 30\\
			\hline
			Trainable parameters $\times10^5$ & 5 & 5.1 & 3.1 & 4.3 & 5.5 & 4.3\\
			\hline

		\end{tabular}
		\caption{\label{table_OD}Optimal CoordConv parameters for different optical thicknesses. For all optical thicknesses, the tuning range used for the hyperparameter search is the same as in Table~\ref{table_cnn}.}
	\end{center}
\end{table}

\subsection{Selection of the best architecture}

\label{sec_selection_architecture}

\begin{figure*}[h]
	\begin{center}
		\includegraphics[width=0.70\textwidth]{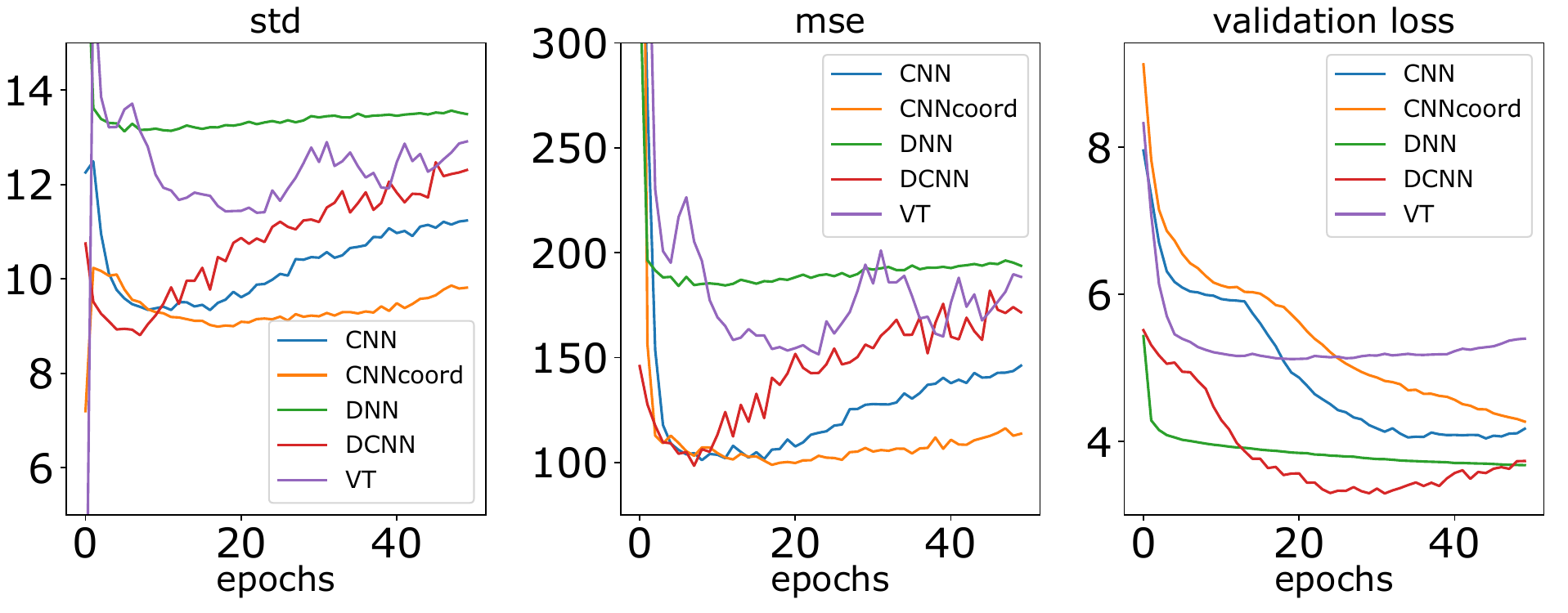}
	\end{center}
	\caption{Performance comparison of different ANN architectures for the $b=4.2$ dataset. Left panel: standard deviation of the predicted data averaged over all $x$ and $y$ positions. Central panel: mean squared error (with respect to the ground truth). Right panel: validation loss.}
	\label{fig_anncomp}
\end{figure*}

We could select the optimal architecture as the one that minimizes either training loss, validation loss, or prediction MSE. In order to compare the ANN architectures, we considered the $b=4.2$ dataset and trained them for 50 epochs. We then compared the standard deviation of the position estimates, the validation loss and the MSE, which are plotted in \fig{fig_anncomp}. In addition to the architectures mentioned in the manuscript, we also tested here a Vision Transformer (VT) ANN~\cite{dosovitskiy2020image_2} with 8 transformer layers, each of these layers including a multi-head self-attention mechanism with 4 parallel attention heads, a feed-forward network with 128 hidden layer units and a GELU activation function, and  normalization layer. The input images are divided into 4$\times$4 patches, which leads to the feature space of a dimension of 64. The target coordinates are encoded using the same 2-hot encoding scheme as in the other architectures, with each coordinate represented as a 64-dimensional vector.

As can be seen from \fig{fig_anncomp}, the CNN and the CoordConv architectures can reach roughly the same standard deviation, but CoordConv reaches a slightly smaller MSE due to a lower bias. Therefore, we consider it as being the optimal architecture for the task. However, we need to mention that we did not perform hyperparameter optimization with VT in the same way as we did for other architectures, as the search space is much larger than for the other architectures; therefore, we cannot exclude that better performances could be obtained by finding better hyperparameters.

\subsection{Bias correction}

\label{sec_bias_correction}

\begin{figure*}[ht]
	\begin{center}
		\includegraphics[width=\textwidth]{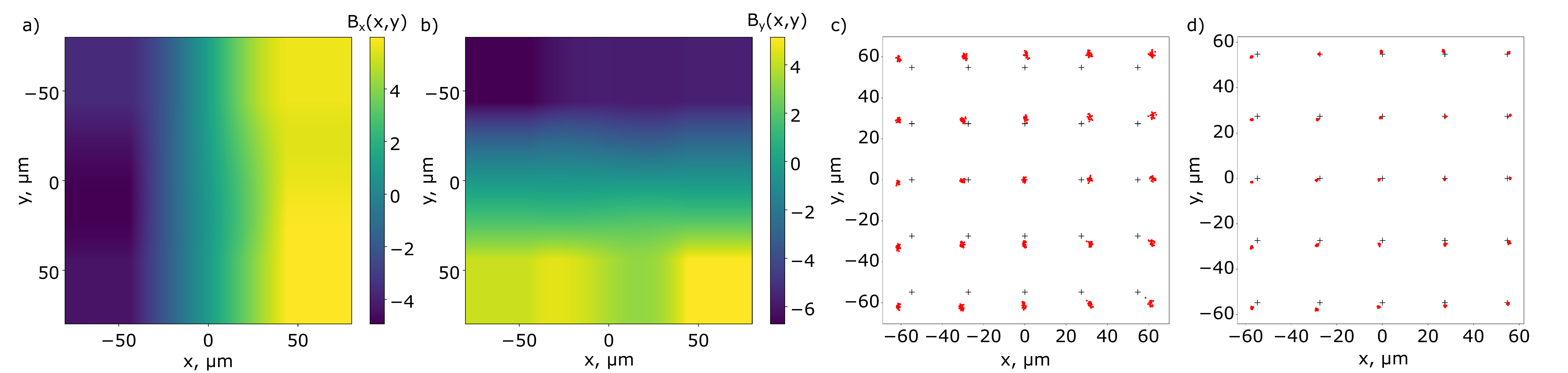}
	\end{center}
	\caption{(a),(b) Bias correction functions in $x$ and $y$. (c) Predicted target position values (red dots) for $b= 2.5$ before and (d) after bias correction. The black crosses in (c) and (d) denote true positions.}
	\label{fig_bias}
\end{figure*}

As can be seen from Fig.~3 of the manuscript, all ANN architectures develop a bias. To compare the ANN predictions to the unbiased Cramér-Rao bound, we need to correct for this bias. For this purpose, we construct a function that characterizes the dependence of the bias on the target position. Initially, we extract a set of patterns from the test dataset, and apply the same augmentation procedure as employed during the training phase. We then pass this set through the trained network and calculate the average difference between the predicted and real values. After that, we fit a 2D spline function $B(x,y)$ to these values, and further use it to correct the predicted coordinates. An example of the fitted $B(x,y)$ (for x and y coordinates) and the effect of the bias correction on the ANN predictions is shown in \fig{fig_bias}.

\subsection{Influence of the size of the test set}

\label{sec_size_dataset}

\begin{figure*}[ht]
	\begin{center}
		\includegraphics[width=0.9\textwidth]{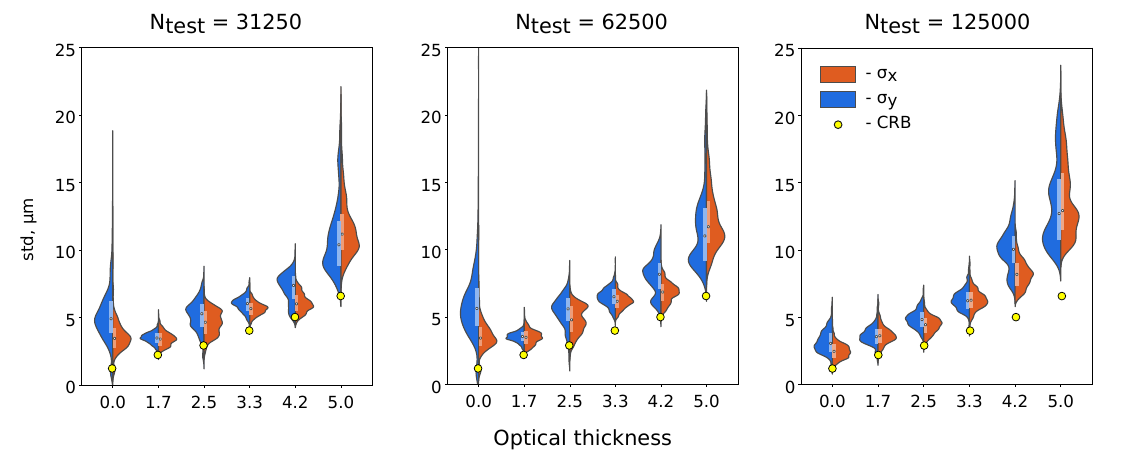}
	\end{center}
	\caption{Test sample size dependence of the ANN (CoordConv) performance. }
	\label{samplesize}
\end{figure*}

From Fig.~3(b) of the manuscript, it seems that the Cram\'er-Rao bound can sometimes be overpassed, for some ANN realizations and some target positions. We attribute this effect to the finite sample size of the test set. In \fig{samplesize}, we plot the ANN prediction histograms while increasing the size of the test set: we first use a quarter of the test data, a half, and finally the full test set. As can be seen, increasing the size of the test set reduces the number of configurations that overpass the bound, which indicates that the finite sample size of the test set is likely to explain the few occurrences in which the Cram\'er-Rao bound seems to be overpassed.

\subsection{Target position dependence}

\label{sec_position_dependence}

\begin{figure*}[ht]
	\begin{center}
		\includegraphics[width=0.8\textwidth]{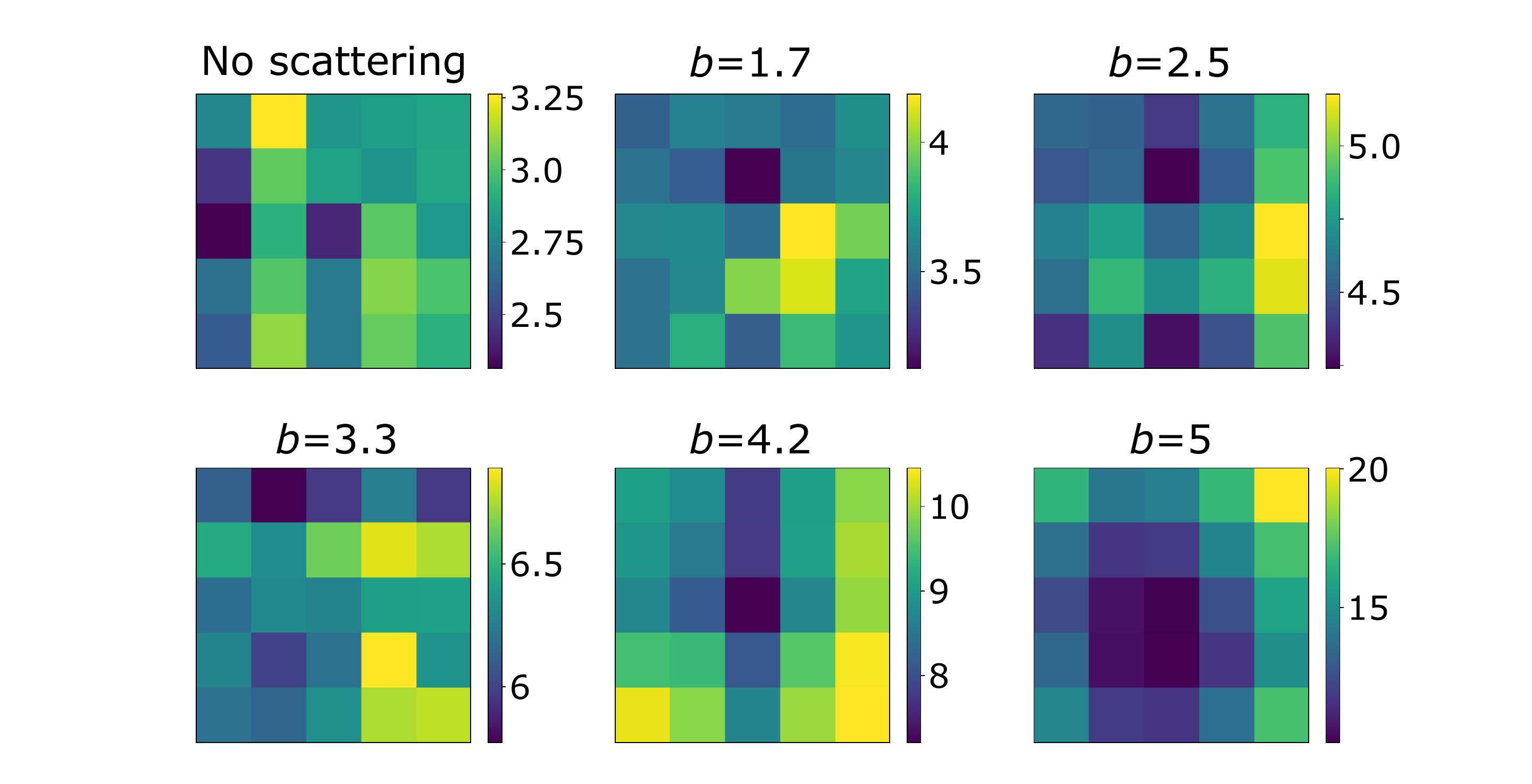}
	\end{center}
	\caption{Dependence of the ANN uncertainty (average for x and y) on the target position averaged over 25 ANN random initializations. }
	\label{targe_pos}
\end{figure*}

In \fig{targe_pos}, we plot the dependence of the average (over 25 initializations, as well as over x and y coordinates) ANN uncertainty on the target position. As can be seen from this plot, the average uncertainty does not depend on the target position for the weakly-scattering samples. When the optical thickness is increased, however, the reduction of the ANN precision does not happen uniformly over all target positions. We indeed observe that the central positions are easier to predict for the network, as compared to the ones located in the corners of the field of view. We attribute this effect to the fact that the corner positions are less connected to the output of the network, which makes it harder for the information to flow from these areas to the output. 

\section{Model generalizability}
	
	\subsection{Influence of the scattering strength}
	
	\label{sec_generalizability_scattering}

\begin{figure*}[ht]
		\begin{center}
			\includegraphics[width=0.8\textwidth]{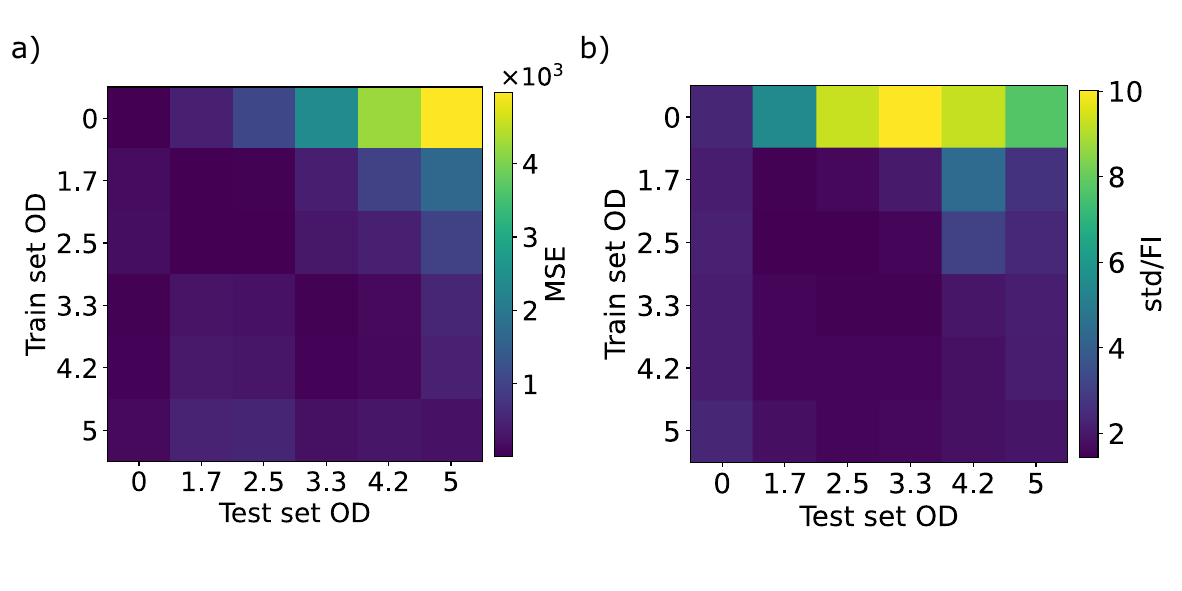}
		\end{center}
		\caption{(a) Mean squared error and (b) standard deviation normalized by the value of the Cramér-Rao bound of the models trained on one optical thickness and tested on another.}
		\label{fig:od_crosscheck}
	\end{figure*}

	ANNs can struggle to generalize to conditions that are not represented in the training dataset. In our experiments, we trained multiple models using experimental data measured for different scattering strengths. It is interesting to assess whether these ANNs can be used to precisely estimate the target position for unseen scattering densities. In \fig{fig:od_crosscheck}, we present cross-validation results, where models trained at a specific optical thickness are tested on data from different densities. We observe that models trained on stronger scattering datasets perform relatively well when tested in weakly scattering conditions, while models trained on weakly scattering datasets becomes much more imprecise when tested in strongly scattering conditions. In order to generalize, it is thus preferable to train the model in the worst-case scenario with regards to the scattering strength.
	
	\subsection{Influence of the object size and shape}
	
	\label{sec_generalizability_size}

	The size of the object has an influence on the localization accuracy. For an object with sharp edges, as in our experiments, information about the target position is only created at the edges of the object \cite{hupfl_continuity_2024_2}. In particular, for a square object of side length $a$, in an ideal free-space imaging configuration and for Gaussian noise statistics, the Fisher information would scale with $a$ and the Cramér-Rao bound would scale with $1/\sqrt{a}$. To verify if a similar behavior can be observed in our experiments (in the presence of a scattering medium), we considered a scattering medium of optical thickness $b=2.5$ and we performed measurements on three different objects of different sizes ($2\times2$ DMD pixels, $5\times5$ DMD pixels and $7\times7$ DMD pixels). From these measurements, we calculated a Cramér-Rao bound of $2.9$\,\textmu m for the biggest object, $3.1$\,\textmu m for the medium-sized object and $5.5$\,\textmu m for the smallest object. This shows that a bigger object indeed produces a larger amount of information, and therefore yields a smaller Cramér-Rao bound. However, the scaling in $1/\sqrt{a}$ is only approximately recovered, indicating that other effects due to the presence of the scattering medium could play a role (such as the complex noise statistics or possible diffraction effects). 
	
	As reported in the manuscript, the observed precision of the ANN is, on average, always higher than the Cramér-Rao bound. Here, we observed a value of $5.2$\,\textmu m for the biggest object, $5.5$\,\textmu m for the medium-sized object and $20.7$\,\textmu m for the smallest object. While the ANN precision stays relatively close to the bound for the biggest object and for the medium-sized object, it becomes significantly higher than the bound for the smallest object. This confirms that, for less favorable signal-to-noise conditions (either due to a smaller object size or to a larger optical thickness), the bound becomes harder to reach. 
	
	We also perform similar measurements on a cross-shaped object, covering $8\times8$ DMD pixels with a line thickness of $3$ pixels. We obtained a Cramér-Rao bound of $3.0$\,\textmu m and an ANN precision of  $5.4$\,\textmu m, very close to what is observed for square objects of similar size. 
	
\begin{figure*}[ht]
	\begin{center}
		\includegraphics[width=0.8\textwidth]{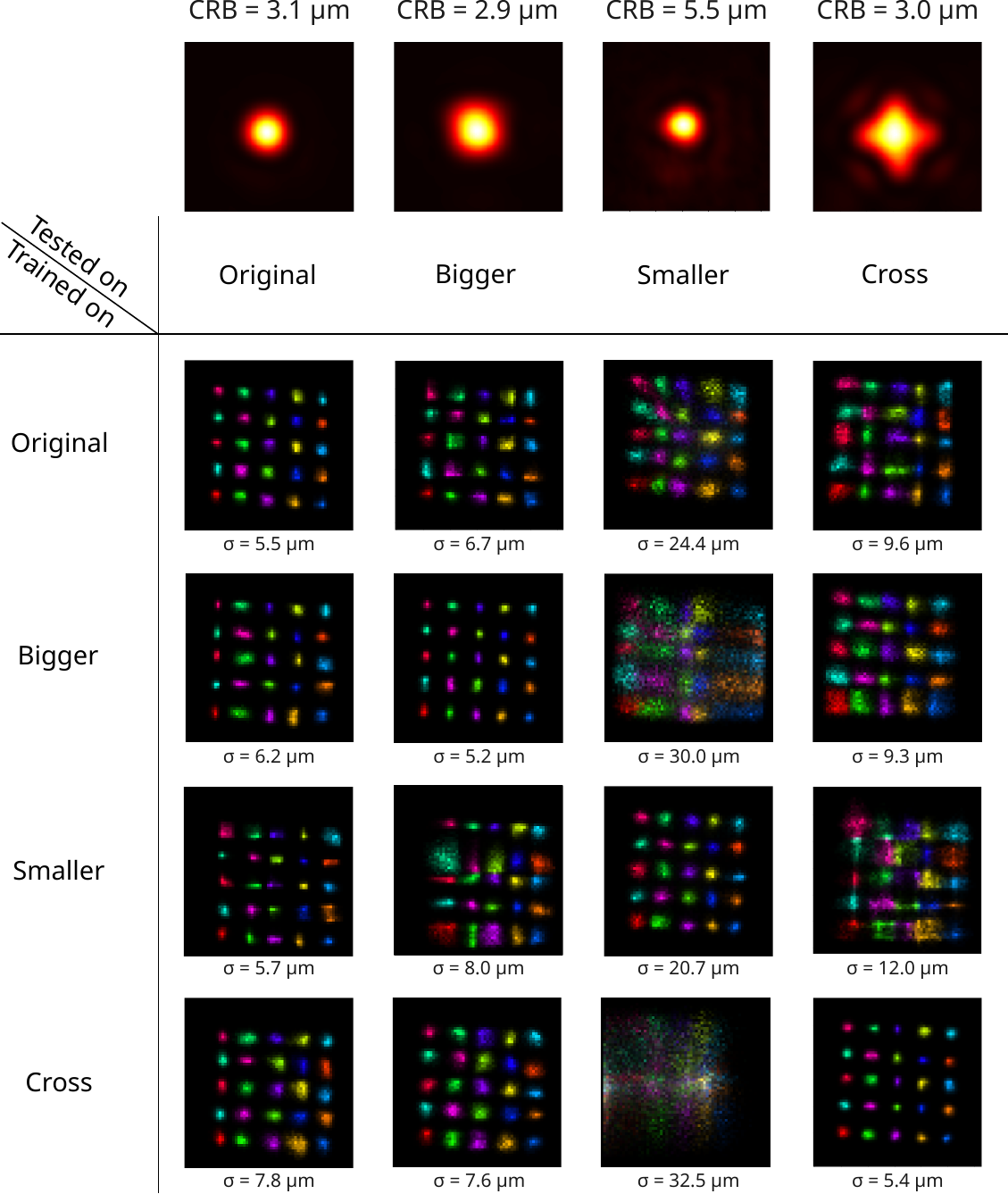}
	\end{center}
	\caption{Standard deviation $\sigma$ of the models trained on one object size and tested on another. The original object is the one presented in the manuscript (composed of $5\times5$ DMD pixels), the bigger object is composed of $7\times7$ DMD pixels, the smaller object is composed of $2\times2$ DMD pixels, and the cross covers $8\times8$ DMD pixels (with a line thickness of $3$ pixels). Training and testing are performed with an optical thickness $b=2.5$. Images of the objects (top row) correspond to the ballistic contributions that are obtained by averaging over many disorder realizations; they are presented for illustration purposes. Note that, due to slight misalignment between the DMD and the camera, the ground truth positions appear as slightly tilted in this figure. This effect is consistent for the different targets and does not affect the ANN standard deviation estimates.}
	\label{fig:obj_size}
\end{figure*}

	Finally, in addition to the cross-test performed for different scattering strengths (see \fig{fig:od_crosscheck}), we performed a cross-test on different object sizes and shapes to evaluate the ability of the ANN to generalize. The results are shown in \fig{fig:obj_size}. For objects with comparable amount of information (1st, 2nd and 4th rows and columns), the ANN is able to generalize relatively well when trained with a different object sizes and shapes, with an increase of the standard deviation ranging from 13\% to 78\% when the ANN is trained and tested with a different object size or shape (depending on the geometrical similarities between the considered objects). In all cases involving the smallest object (3rd raw and column), the situation changes as the amount of information in the testing data is significantly smaller. Then, training the ANN on different object sizes and shapes leads to poor results (3rd column). Note that, when trained on the smallest object and evaluated using images with more information (3rd line), the standard deviation decreases: the larger amount of information available in the data then compensates the penalty induced by training and testing on a different object size and shape.

\end{document}